# Electrically tunable interfacial thermal conduction via electronic structure engineering in Au/$Bi_{1-x}Sb_x$ topological insulators

Min Young Kim[1,2]*, Joon Sang Kang[3], Jangwoo Ha[4], Hyungyu Jin[4], Sandy A. Ekahana[5], Pratik Saud[5], Chris Jozwiak[6], Eli Rotenberg[6], Aaron Bostwick[6], Jyoti Katoch[5], Joseph P. Heremans[1,7,8]*

*[1]Department of Mechanical and Aerospace Engineering, The Ohio State University, Columbus, OH 43210, USA.*

*[2]Electronic and Hybrid Materials Research Center, Korea Institute of Science and Technology (KIST), Seoul 02792, South Korea.*

*[3]Department of Mechanical Engineering, Korea Advanced Institute of Science and Technology (KAIST), Daejeon 34141, South Korea.*

*[4]Department of Mechanical Engineering, Pohang University of Science and Technology (POSTECH), Pohang 37673, South Korea.*

*[5]Department of Physics, Carnegie Mellon University, Pittsburgh, Pennsylvania 15213, USA.*

*[6]Advanced Light Source, E. O. Lawrence Berkeley National Laboratory, Berkeley, California 94720, USA.*

*[7]Department of Physics, The Ohio State University, Columbus, OH 43210, USA.*

*[8]Department of Materials Science and Engineering, The Ohio State University, Columbus, OH 43210, USA.*

*Corresponding author. Email: minykim@kist.re.kr (M.Y.K.); heremans.1@osu.edu (J.P.H.)

## Abstract

This work provides direct experimental evidence for the role of topological interface states in thermal conduction across a metal/topological insulator junction. It also shows that this conduction can be reversibly modulated by electrical current injection, offering a new approach toward active control of heat flow at solid-state interfaces. Specifically, the interfacial thermal conductance of Au/$Bi_{89}Sb_{11}$ and Au/$Bi_{87}Sb_{13}$ junctions demonstrates distinct temperature- and bias-dependent behavior. Both responses are attributed to carrier redistribution between topological interface and bulk band states, driven thermally by Fermi-Dirac broadening and electrically by quasi-Fermi-level shifts and WKB tunneling into nearby bulk bands. Control experiments using trivial semimetals and insulating interlayers further confirm the topological specificity of the effect. Such electrically tunable interfacial heat conduction positions interface electronic structure engineering as a promising route for active thermal management. In doing so, it lays the groundwork for a mechanically robust alternative to conventional structure-driven thermal control compatible with increasingly dense, high-power solid-state devices.

## Introduction

From the smartphones in our pockets to the quantum computers poised to redefine information processing, modern electronic systems are packing ever-increasing power densities and integration levels. Such compact architectures often overheat, which is responsible for approximately 55% of electronic failures (*1, 2*). Efficient dissipation of such excess heat is often limited by solid-solid interfaces, making interfacial thermal transport a key determinant of device performance and reliability. The major energy carriers across these interfaces are phonons and electrons, and their interfacial transport physics has therefore been extensively investigated (*3-7*). For phonon-dominated transport, interfacial heat flow has been tailored through engineering of vibrational spectral matching and extrinsic interfacial structures, both of which reduce interfacial thermal resistance by promoting phonon transmission across boundaries (*8-13*). Similar interface-engineering strategies have also been employed in electron-mediated systems, particularly metal-semiconductor heterostructures, where strengthened interfacial electron-phonon coupling facilitates cross-interface energy transfer. Direct electron transmission has likewise been improved by modulating near-interface electronic band profiles between dissimilar materials (*14-18*). These microscopic insights have enabled substantial enhancement of the interfacial thermal conductance ($G$). Yet, practical thermal management increasingly requires active regulation of interfacial heat flow in response to changing operational demands, rather than simply maximizing $G$ for passive cooling.

Active thermal management seeks on-demand heat dissipation tailored to real-time system requirements through functional components such as thermal switches, diodes, and transistors (*19-21*). Such functionalities are typically realized by manipulating heat-carrying excitations under external stimuli, including electric (*22-26*) or magnetic fields (*27, 28*), strain (*29, 30*), temperature (*31*) or pressure (*32*). Most demonstrations, however, have focused on altering bulk thermal conductivity of constituent materials through structural phase transitions, whereas direct control of $G$ has received considerably less attention despite its practical relevance. A few studies have begun to address this gap, pointing to the potential for tunable $G$ via mechanisms such as moiré superlattice-induced phonon scattering (*33*) and electric-field-controlled interfacial binding interactions in molecular junctions (*23, 34*). These advances established the viability of $G$ modulation, but extending this capability to scalable and reproducible solid-state material platforms remains an important challenge. Equally important is the absence of clear design rules for identifying interfaces capable of large changes in $G$, as progress in the field has largely relied on case-specific investigations.

Here, we pursue a distinct strategy by exploiting topological surface states (TSS) as heat-conduction channels. TSS are conducting states that reside on the surface of topological insulators (TIs), protected by time-reversal symmetry and strong spin-orbit coupling. The hallmarks of TSS are spin-momentum locking and Dirac-like dispersion, which suppress electron backscattering and support highly mobile charge carriers (*35-37*). These characteristics enable exceptional surface electrical conduction (*38, 39*). According to the Wiedemann-Franz law, such electronic channels are therefore expected to support efficient thermal transport as well. Indeed, enhanced in-plane electrical and thermal conduction is observed in ultrathin TI films, where the contribution of TSS to charge and heat transport is maximized by the large surface-to-volume ratio (*40-44*). In this context, $Bi_{1-x}Sb_x$ alloys are particularly attractive for exploring TSS-mediated thermal transport. They are among the best-characterized 3D TIs and host a fivefold-degenerate surface-state manifold at the Fermi level ($E_F$), giving rise to a high density of states (DOS) and remarkably high surface conductance (*45-50*). Importantly, owing to the evolution of the bulk band structure with Sb concentration ($x$), $Bi_{1-x}Sb_x$ enters a topologically non-trivial insulating phase for $x > 8.6\%$, where robust TSS are stabilized and progressively dominate low-energy surface transport (*51, 52*).

To harness TSS for interfacial heat transport, we first consider metal-TI junctions, where topological interface states (TIS) emerge when a metal is brought into contact with a TI. These TIS are expected to largely retain the Dirac spectrum under sufficiently weak hybridization with metal electrons (*53*). When a $Bi_{1-x}Sb_x$ TI is interfaced with Au, the presence of TIS should therefore facilitate more efficient heat transfer across the junction than in junctions formed with trivial semiconducting $Bi_{1-x}Sb_x$ (**Fig. 1**, **A** and **B**). As shown in **Fig. 1C**, a Schottky barrier forms at the Au-$Bi_{1-x}Sb_x$ interface in the semiconducting phase as a consequence of band alignment and Fermi-level equilibration (*54*). However, the formation of TIS enriches the electronic spectrum near the interface, thereby increasing the density of electronic states available for transmission across the junction (**Fig. 1D**). Given that electrons are the primary energy carriers in the Au/$Bi_{1-x}Sb_x$ system, this TIS-mediated electronic pathway may meaningfully contribute to interfacial thermal conduction. In this framework, we electrically modulate $G$ at the Au-$Bi_{1-x}Sb_x$ interface via current injection. The applied bias current induces a non-equilibrium electrochemical potential difference of approximately $qV$ ($V$: junction voltage) across the interface, which in turn alters the occupation of TIS and the density of available states participating in transport. Indeed, the resulting bias-dependent variations in $G$ provide direct experimental evidence that TIS can serve as reconfigurable channels for interfacial thermal transport.

## Results

### *Crystal quality and TSS in $Bi_{1-x}Sb_x$*

We employ three $Bi_{1-x}Sb_x$ single crystals previously reported in Ref. (*55*): A2 ($x$ = 3%), A6 ($x$ = 11%), and A7 ($x$ = 13%). A2 is a trivial semimetal without TSS, whereas A6 and A7 lie within the topological insulating regime ($x$ > 8.6%). For convenience, we now refer to these samples as $Bi_{97}Sb_3$, $Bi_{89}Sb_{11}$, and $Bi_{87}Sb_{13}$, respectively. As shown in **Fig. 2A**, all crystals exhibit high carrier mobilities ($\mu_{[001]}$), approaching $10^6$ $cm^2 \cdot V^{-1} \cdot s^{-1}$ at low temperatures and exceeding $10^4$ $cm^2 \cdot V^{-1} \cdot s^{-1}$ at room temperature, indicative of excellent crystalline quality. As shown in **Fig. 2B**, the carrier concentration ($n$) of $Bi_{97}Sb_3$ saturates near $10^{17}$ $cm^{-3}$ below 30 K, in agreement with its semimetallic nature. In contrast, the insulating bulk character of $Bi_{89}Sb_{11}$ and $Bi_{87}Sb_{13}$ leads to progressively lower $n$ at low temperatures. Arrhenius analysis (**Fig. 2C**) yields band gaps ($E_g$) of approximately 31.1 meV and 20.9 meV, respectively, with the latter likely reduced by H-band involvement at higher $x$ (*56, 57*). X-ray diffraction measurements reported in Ref. (*55*) further confirm the compositional uniformity, revealing less than 1% variation in $x$ over 2-cm crystal segments. In line with Ref. (*58*), the thermal conductivity ($\kappa_{[001]}$) increases with temperature ($T$) across all compositions, while its magnitude decreases with increasing $x$, presumably due to enhanced phonon scattering from additional Sb atoms (**Fig. 2D**).

The presence of TSS in the synthesized $Bi_{1-x}Sb_x$ crystals was examined by angle-resolved photoemission spectroscopy (ARPES) measurements performed on a freshly cleaved hexagonal (001) surface. **Fig. 3A** shows a schematic of the bulk Brillouin Zone (BZ) of elemental Bi, where the bulk Fermi surface consists of electron and hole pockets centered near the L and T points, respectively (*51, 52*). The projected surface BZ contains the high-symmetry points $\overline{\Gamma}$, $\overline{\mathrm{M}}$, and $\overline{\mathrm{K}}$. Here, we focus on the $\overline{\Gamma}-\overline{\mathrm{M}}$ direction, along which the TSS continuously spans from the bulk T valence band continuum near $\overline{\Gamma}$ toward the L conduction band edge near $\overline{\mathrm{M}}$. **Fig. 3** (**B** and **C**) shows sharp and intense Fermi surface features of $Bi_{89}Sb_{11}$ at 80 K and 300 K, respectively, consisting of highly anisotropic ellipsoidal contours. The corresponding band dispersions are shown in **Fig. 3** (**D** and **E**), which reveal two well-defined surface bands, denoted S1 and S2. Consistent with prior ARPES studies (*45, 47*), S1 emerges linearly near $\overline{\Gamma}$ and evolves into an electron pocket close to $\overline{\mathrm{M}}$ (hereafter denoted S1' for clarity), whereas S2 forms a hole pocket around $\overline{\Gamma}$. Together, these two surface bands cross $E_F$ five times, which preserves the odd number of crossings required for a topologically non-trivial phase and thereby confirms the existence of TSS in $Bi_{89}Sb_{11}$.

Upon increasing temperature from 80 K to 300 K, the spectral linewidth broadens noticeably due to enhanced electron scattering, with the characteristic surface-state dispersion more clearly resolved at 300 K. At the same time, the T valence band continuum shifts slightly upward at 300 K relative to 80 K, accompanied by a reduced bulk band gap (*51*). The S1 and S2 surface states, which emerge from this continuum, follow the same upward shift while retaining their overall dispersion topology. A quantitative analysis of the S1 and S2 dispersions at 300 K yields a Fermi velocity of $v_F = (4.53\pm0.05)\times10^5$ m·s$^{-1}$ for the linear S1 branch. The S1' and S2 pockets exhibit effective masses of $(1.73\pm0.01)m_e$ and $(3.14\pm0.05)m_e$, respectively. Here, S1 near $\overline{\Gamma}$ was approximated as a linear band, whereas both the hole and electron pockets were treated as parabolic dispersions. In addition, the band-edge separation, defined as $\Delta E = E_C - E_V$, is determined to be 24 meV. These parameters will be revisited below to elucidate the origin of the observed modulation of $G$. Although the ARPES measurements were performed only on $Bi_{89}Sb_{11}$, prior ARPES studies report that S1 and S2 energy positions vary smoothly with $x$ across the topological transition (*47*), supporting the expectation that $Bi_{87}Sb_{13}$ hosts a similar surface-state configuration.

### *Signatures of TIS in thermal conductance at Au-$Bi_{1-x}Sb_x$ interfaces*

In this work, we employed frequency-domain thermoreflectance (FDTR) (*59-62*), a highly sensitive pump-probe technique for measuring $G$. We deposited a 100 nm-thick Au layer on the cleaved $Bi_{1-x}Sb_x$ (001) surface, and measured the interfacial thermal conductance along the [001] trigonal axis ($G_{[001]}$). In FDTR, a modulated pump laser periodically heats the sample surface, while a synchronized probe laser monitors temperature-induced reflectance changes of Au transducer at each modulation frequency ($f$). The resulting phase lag ($\varphi$) between the thermal excitation and the surface response serves as the primary observable, from which $G_{[001]}$ is extracted by fitting to a thermal diffusion model (**Fig. 4A**). To validate the reliability of our FDTR measurements, we assessed the repeatability of the phase spectra through eight consecutive frequency sweeps under thermally stabilized conditions. The relative standard deviation of these signals remained below 0.5% across the entire frequency range, confirming the excellent stability of our FDTR system (**fig. S4**). **Fig. 4** (**B** and **C**) shows the resulting $T$-dependent $G_{[001]}$ of various Au/$Bi_{1-x}Sb_x$ junctions. At each temperature, three independent frequency sweeps were acquired, and the $G_{[001]}$ values were defined as the average of the three measurements. The variation among three independent measurements did not exceed 1% over the full $T$ range, demonstrating the high measurement stability. While the absolute value of $G_{[001]}$ carries a systematic uncertainty inherent to geometric factors and thermal input parameters, this uncertainty affects all measurements equally and therefore largely cancels in relative changes of $G_{[001]}$. The relative trends are instead governed by the precision of repeated FDTR measurements and instrumental resolution, which yield a substantially smaller uncertainty (~1%). Further details of the experimental setup and measurement repeatability analysis are provided in Materials and Methods.

Non-monotonic variations in $G_{[001]}$ with $T$ are observed for both Au/$Bi_{89}Sb_{11}$ and Au/$Bi_{87}Sb_{13}$ samples (**Fig. 4B**). Specifically, $G_{[001]}$ increases with $T$ at low temperatures, drops sharply around 120 K, and then rises rapidly above 220 K. The nearly identical behavior observed in two independent TI samples suggests the presence of a common transport mechanism, the origin of which is discussed below. In contrast, both control samples—Au/$Bi_{97}Sb_3$ and Au/10 nm-thick $Al_2O_3$/$Bi_{87}Sb_{13}$—display a smooth, monotonic increase in $G_{[001]}$ over the entire $T$ range without any discernible anomaly (**Fig. 4C**). The magnitude of Au/$Bi_{97}Sb_3$ remains higher than that of the Au/TI samples, consistent with the metallic nature of $Bi_{97}Sb_3$, while the disappearance of the non-monotonic feature upon $Al_2O_3$ insertion points to the involvement of an electronically mediated transport channel at the interface. The convergence of these two independent control strategies, the absence of TSS in $Bi_{97}Sb_3$ and electronic decoupling via $Al_2O_3$, on the same anomaly-free $T$ dependence argues against a coincidental origin. This interpretation is further supported by the smooth $T$ dependences of the bulk transport properties $\mu_{[001]}$, $n$, and $\kappa_{[001]}$ (**Fig. 2**), which rule out a bulk origin for the anomalous $G_{[001]}$ behavior.

To elucidate the underlying mechanism of the observed $T$ dependence of $G_{[001]}$, we analyzed interfacial thermal transport across the Au-$Bi_{1-x}Sb_x$ interface using an electronic formulation of the diffusive mismatch model (eDMM) (*3, 63*), in which interfacial heat transfer is treated as electron-mediated, with momentum randomized but energy conserved during transmission. Within this framework, we treat electron-electron and electron-phonon channels as the dominant contributors to the total interfacial thermal conductance ($G_{tot}$), given the established role of electrons as primary heat carriers in metal-mediated interfacial transport; how well this picture accounts for the observed $T$-dependent behavior of $G_{[001]}$ is discussed below, including the plausibility of a residual phonon-phonon contribution. For the case without TIS (**Fig. 5A**), interfacial heat transfer is governed by electron-electron coupling between Au electrons and $Bi_{1-x}Sb_x$ bulk electrons, represented by the conductance term $G_{ee1}$. The transferred energy then dissipates into the $Bi_{1-x}Sb_x$ bulk phonon reservoir through bulk electron-phonon coupling ($G_{ep1}$). In the limit of strong electron-phonon interactions in the bulk (i.e., $G_{ep1} \gg G_{ee1}$), $G_{tot}$ is dominated by $G_{ee1}$ and can therefore be approximated as $G_{tot} \approx G_{ee1}$. For the parabolic bulk L-band electrons, $G_{ee1}$ scales as (see Materials and Methods for the detailed derivation):

$$G_{ee1} \approx \frac{k_B{}^2 T}{\hbar^3} m^*_{DOS} \left| E_{C(V)} - E_F \right| \quad (1)$$

where $m_{DOS}{}^*$ is the effective DOS mass.

By contrast, when TIS are present (**Fig. 5B**), thermal transport involves additional energy-transfer pathways associated with the interface carriers. Thermal energy is first transferred from Au electrons to TIS through electron-electron coupling ($G_{ee2}$). The injected energy is subsequently released to bulk phonons through two parallel channels: (1) indirectly, via TIS-to-bulk electron coupling ($G_{ee3}$), followed by electron-phonon scattering ($G_{ep1}$); and (2) directly, via TIS-to-phonon coupling ($G_{ep2}$). Here, the overall energy relaxation from TIS to the $Bi_{1-x}Sb_x$ bulk ($G_{bulk}$) is expected to remain much faster than the interfacial electron-electron coupling ($G_{ee2}$), since TIS are intrinsically part of the same crystal as the bulk reservoirs rather than a separate layer. $G_{tot}$ can therefore be approximated by $G_{ee2}$. Because the functional form of $G_{ee2}$ depends on the interface band dispersion, separate expressions are obtained for the linear S1 branch ($G_{ee2,S1}$), the parabolic electron pocket near $\overline{\mathrm{M}}$ ($G_{ee2,S1'}$) and the hole pocket ($G_{ee2,S2}$) discussed above:

For the linear band S1,

$$G_{\mathrm{ee2,S1}} \approx \frac{k_{\mathrm{B}}{}^{2}T}{48\hbar^{3}v_{\mathrm{F}}{}^{2}}\left(E_{0}-E_{\mathrm{F}}\right)^{2} \quad (2)$$

where $v_F$ is the Fermi velocity and $E_0$ is referenced to the Dirac point.

For the parabolic dispersions,

$$G_{\mathrm{ee2,S1'}} \approx \frac{k_{\mathrm{B}}{}^{2}T}{2\hbar^{3}} m_{\mathrm{DOS}}^{*}\left(E_{\mathrm{F}}-E_{\mathrm{C}}\right) \quad (3) \quad \text{and} \quad G_{\mathrm{ee2,S2}} \approx \frac{k_{\mathrm{B}}{}^{2}T}{2\hbar^{3}} m_{\mathrm{DOS}}^{*}\left(E_{\mathrm{V}}-E_{\mathrm{F}}\right) \quad (4)$$

To establish the hierarchy among these transport channels, we estimated the electron-electron thermal conductances at 80 K using the parameters extracted from the literature (*64*) together with the ARPES-derived $v_F$, $m^*$ and $\Delta E$. The use of these bare-surface parameters is supported by ARPES studies on the related TI $Bi_2Se_3$, which show that the topological surface dispersion persists under Au contact, at least at the coverages probed (*65*). The estimated conductances are $G_{ee1} \approx 0.26$, $G_{ee2,S1} \approx 1.3$, $G_{ee2,S1'} \approx 14$, and $G_{ee2,S2} \approx 8.3$ MW·m$^{-2}$·K$^{-1}$ for the respective channels. Notably, the S1' and S2 channels are over an order of magnitude more conductive than the L-band ($G_{ee1}$) and the linear S1 branch ($G_{ee2,S1}$). These results indicate that the parabolic interface pockets serve as the primary channels for interfacial thermal transport.

Building on this picture, the non-monotonic $T$ dependence of $G_{[001]}$ can be understood as a thermally driven crossover in the dominant interfacial transmission pathway. At 80 K, the thermal broadening ($\sim 3k_BT = 21$ meV) is smaller than the energy separation between the S1' and S2 pockets ($\Delta E = 24$ meV), which lie within the bulk L-band gap, such that interfacial thermal transport proceeds primarily through TIS. As $T$ increases toward 120 K, the expanding thermal window encompasses a larger fraction of the TIS DOS, increasing the number of states available for electron transmission and driving the initial rise in $G_{[001]}$. Around 120 K, however, $3k_BT$ becomes comparable to the bulk band-gap scale (~31.1 meV, **Fig. 2C**), such that the Fermi-Dirac tail extends appreciably toward the L-band edge, rendering these states thermally accessible (~18% occupation). Because the 3D bulk DOS near the band edge scales as $\sqrt{(E-E_C)}$, even this modest thermal broadening produces a rapid onset of bulk states available for additional interband scattering (*66*). This threshold increase in bulk DOS drives a redistribution of electronic occupation away from TIS toward bulk channels, suppressing the effective TIS-mediated transmission—even though the bulk channel itself ($G_{ee1} \approx 0.26$ MW·m$^{-2}$·K$^{-1}$) is far less conductive than the TIS channels. The apparent sharpness of the $G_{[001]}$ suppression near 120 K (**Fig. 4B**) thus reflects a shift from TIS-mediated conduction toward bulk-scattering-limited transport. In the intermediate temperature regime (120-220 K), these competing effects result in only a gradual net increase in $G_{[001]}$. At still higher temperatures, the T-band heavy-hole states shift closer to $E_F$ and become increasingly susceptible to thermal excitation. Given their large effective mass and DOS (*64, 67*), the onset of T-band involvement drives a rapid increase in $G_{[001]}$ above 220 K, accompanied by the increasing metallic character reflected in the $T$-dependent resistivity (**fig. S7**). While this electron-channel picture accounts for the principal features of the $T$-dependent $G_{[001]}$, a residual phonon-phonon contribution cannot be rigorously excluded. We note, however, that neither Au nor the lattice contribution to $\kappa_{[001]}$ ($\kappa_{lat}$, **fig. S10**) exhibits any anomaly over the investigated temperature range, making such a contribution unlikely to underlie the observed non-monotonic behavior.

Having identified TIS as the primary contributors to interfacial thermal transport, we next examine the electrical tunability of $G_{[001]}$ below 120 K, where the influence of adjacent bulk bands remains minimal. As shown in **Fig. 6**, $G_{[001]}$ under applied bias current density ($j$) displays pronounced symmetric behavior with respect to current polarity in both the $Au/Bi_{89}Sb_{11}$ and $Au/Bi_{87}Sb_{13}$ samples. Specifically, $G_{[001]}$ increases sharply at low $|j|$, reaches a maximum, and subsequently decreases at higher $|j|$. Furthermore, the peak current density shifts systematically toward lower values with increasing *T*.

The microscopic origin of this conductance modulation can be understood in terms of a current-induced quasi-Fermi-level shift at the $Au/Bi_{1-x}Sb_x$ interface. The applied bias current induces a non-equilibrium electrochemical potential difference of approximately $qV$ across the interface, whose sign reverses with the polarity of $j$ (**Fig. 1**). Positive current shifts the quasi-Fermi level toward the electron side of the TIS spectrum, preferentially weighting the S1' pocket, and negative current toward the hole side, preferentially weighting S2, although both channels contribute at all polarities given the small magnitude of $qV$ (~0.5 meV). For positive $j$, Eq. (3) predicts that $G_{ee2,S1'}$ increases as the separation between $E_C$ and $E_F$ becomes larger. Consistent with this expectation, $G_{[001]}$ at 80 K exhibits a nearly linear increase with increasing $j$ up to 2.79 kA·m$^{-2}$, corresponding to $qV = 497$ μeV, which yields $\Delta G_{ee2,S1'} = 0.34$ MW·m$^{-2}$·K$^{-1}$, comparable in magnitude to the observed $\Delta G_{[001]} = 0.118 \pm 0.028$ MW·m$^{-2}$·K$^{-1}$. Beyond $j =$ 2.79 kA·m$^{-2}$, however, $G_{[001]}$ decreases sharply despite the continued increase in $qV$, which suggests the development of a conductance-limiting mechanism. Given the similarity to the temperature-induced conductance suppression, where L-band activation was identified as the limiting mechanism, we evaluate the accessibility of the L-band states through the WKB tunneling probability $P \approx \exp(-2WK)$ with $K = \sqrt{2m^* \Delta\Phi}/\hbar$ (*W*: the barrier width and ΔΦ: the barrier height) (*68*). Assuming $W = 2.5$ nm (*46*) and an effective barrier height of $\Delta\Phi = (E_g - \Delta E)/2 - qV$, the estimated *P* reaches 21% at the 80 K conductance peak, indicating that nearby L-band states can no longer be neglected. A comparable activation scale emerges from the temperature-dependent analysis, where thermal broadening renders ~18% of L-band states accessible at 120 K. The convergence of these independent estimates suggest that current-induced quasi-fermi-level shifts and thermal activation represent analogous routes to L-band accessibility, as evidenced by the similar suppression of $G_{[001]}$ observed in both cases. Eq. (4) predicts the corresponding response for negative $j$ through the S2 hole pocket, accounting for the nearly symmetric $G_{[001]}$-$j$ characteristics with respect to current polarity. The conductance peak also shifts toward lower $|j|$ with increasing $T$, as smaller $qV$ is required to access the bulk L-band states. Taken together, these results reveal three key features of $G_{[001]}$: (1) a conductance maximum governed by quasi-Fermi-level tuning of TIS occupation, followed by suppression via L-band activation at higher bias, (2) a nearly symmetric current response arising fom the dual electron and hole TIS channels, and (3) a systematic shift of the conductance peak toward lower $|j|$ with increasing $T$, reflecting the reduced energy barrier to L-band activation at elevated temperatures.

### *Control experiments confirming the topological specificity of $G_{[001]}$ modulation*

To establish that the observed $G_{[001]}$ modulation originates from TIS rather than bulk transport or trivial electronic effects, we performed three sets of control experiments. First, we examined whether current injection modifies bulk transport properties. As shown in **Fig. 7A**, $\mu_{[001]}$ of $Bi_{89}Sb_{11}$ is insensitive to $j$ over the entire current range at 80 K, with most points lying within ±0.5% of the mean value. Likewise, $\kappa_{[001]}$ is essentially independent of $j$ over the same range at 80 K, confirming that bulk carrier and heat transport are unaffected by current injection. The voltage drop across the Au/$Bi_{89}Sb_{11}$ junction follows a linear $I$-$V$ relationship (**fig. S9**), with the differential resistance *dV/dI* showing no appreciable variations between 80 and 103 K, ruling out current-induced changes in overall electrical resistance. Second, we assessed the role of TIS by measuring $G_{[001]}$-$j$ at elevated temperatures, where the influence of TIS occupation is reduced. As shown in **Fig. 7C**, the Au/$Bi_{89}Sb_{11}$ sample maintains nearly constant $G_{[001]}$ values at 155, 205 and 290 K across the full current range, consistent with the diminished TIS contribution at these temperatures. Third, we examined the Au/$Bi_{97}Sb_3$ control sample, which does not host TIS. As shown in **Fig. 7D**, $G_{[001]}$ remains flat across the full $j$ range at 80, 154, and 290 K, confirming that the current-sensitive response requires the presence of TIS.

## Discussion

Our observations establish a new design paradigm for active thermal interfaces, in which the electronic structure of the interface—rather than its geometry or chemistry—serves as the primary control variable for heat transport. This shift in perspective is enabled by the unique properties of topologically protected interface states, which provide highly conductive and electrically reconfigurable channels for interfacial heat flow. The demonstration that TIS occupation can be modulated by current injection to tune $G$ opens a previously unexplored direction in thermal interface engineering. While demonstrated here in Au/$Bi_{1-x}Sb_x$, the underlying principle—that band topology defines the functional thermal response—is not specific to this system and may guide the development of a broader class of topological heterostructures for active thermal management.

The present results also delineate the key factors that currently limit the magnitude of $G$ modulation. In $Bi_{1-x}Sb_x$, the small bulk band gap of only several tens of meV constrains the accessible quasi-Fermi-level tuning range, as bulk band states become active before the full modulation regime is reached. Future improvements may therefore be realized through TIs possessing larger band gaps, which would extend the accessible quasi-Fermi-level tuning range, and enhanced interfacial DOS, which would strengthen the conductance response within that range. Electrostatic gating offers an additional route for independently shifting the Fermi level position within the band gap, complementing the material-based strategies above. Equally important is the optimization of interface quality to minimize parasitic scattering and improve carrier transmission. Together with such advances in materials and interface engineering, the even-in-field nature of the response may further enable multilayer architectures incorporating repeated Au/TI interfaces. These considerations outline a concrete path toward stronger $G$ modulation and, ultimately, toward practical active heat management based on electronically programmable interfacial transport.

## Materials and Methods

### *Sample preparation and characterization*

All $Bi_{97}Sb_3$, $Bi_{89}Sb_{11}$, and $Bi_{87}Sb_{13}$ single crystals used in this study correspond to samples A2, A6, and A7, respectively, from Ref. (*55*), where they were synthesized by the travelling molten zone method. Detailed descriptions of the crystal growth procedure and bulk transport characterization are provided in Ref. (*55*).

The crystals were cleaved along their natural cleavage plane to expose fresh (001) surface. The surface orientation was confirmed by X-ray diffraction measurements, which showed only (00*l*) reflections (**fig. S1**). A 100 nm-thick Au transducer layer was subsequently deposited on the cleaved surface using e-gun evaporator (CHA Industries) under high vacuum ($\sim 10^{-6}$ torr). Optical images of Au-coated samples are shown in **fig. S2**. Each Au-coated sample was subsequently mounted on a copper heat sink inside a cryogenic vacuum chamber of the FDTR setup.

### *Bulk electrical transport property measurements*

The bulk electrical transport properties, including resistivity and Hall resistivity, were measured using a standard four-point probe configuration. Electrical contacts were made using 25 μm diameter copper wires affixed with silver epoxy. A thin, continuous layer of silver epoxy was applied to the current electrodes to promote uniform current distribution across the sample. Measurements were performed at discrete temperature points, each preceded by a 30 min thermal stabilization period to ensure steady-state conditions. All electrical measurements were carried out using an AC resistance bridge (Model 370, Lake Shore Cryotronics, Inc.) integrated with a Physical Property Measurement System (PPMS, Quantum Design).

The dominant uncertainty in the resistivity measurements arises from geometric factors associated with the finite size of the electrical contacts and alignment tolerances, resulting in an estimated ~10% systematic uncertainty in the absolute resistivity values. This uncertainty does not affect the temperature- or current-dependent trends discussed in the main text.

### *Bulk thermal transport property measurements*

The thermal conductivity and Seebeck coefficient measurements were performed using a steady-state method with differential thermometry (**fig. S10**). Samples were wired using 25 μm copper and constantan leads and mounted on a thermally conductive alumina substrate. A 120 Ω resistive heater attached to the sample surface was used to establish a controlled thermal gradient. The assembly was installed in a custom liquid-nitrogen cryostat system (Lake Shore Cryotronics, Inc.). Measurements were performed after a 30-min thermal stabilization period at each temperature. The temperature difference across the sample and the corresponding Seebeck voltage were measured using a nanovoltmeter (Model 2182A, Keithley). All measurements were performed under high vacuum (~$10^{-6}$ torr) with a gold-plated radiation shield to minimize radiative heat losses.

The thermal conductivity measurements are subject to two primary sources of systematic uncertainty: (1) geometric imprecision in determining the heat-flow path length, and (2) uncertainty in the applied heat flux. The geometric contribution, arising from finite contact sizes, and alignment tolerances, is estimated to be approximately 10%. Uncertainty in the applied heat flux mainly arises from parasitic heat losses through wiring and residual radiation. These losses remain significantly smaller than the measured thermal conductance and become negligible below 200 K. In contrast, Seebeck coefficient measurements are less sensitive to geometric uncertainty because they depend on the ratio of the thermoelectric voltage to temperature gradient. These systematic uncertainties primarily affect the absolute values and have negligible influence on the relative temperature- and current-dependent variations in the main text.

## *ARPES measurements*

Synchrotron-based ARPES measurement were performed at Beamline 7.0.2 (MAESTRO) of Advanced Light Source (ALS), Berkeley, CA, USA at a temperature of 80 K and 300 K and pressure less than $5\times10^{-11}$ torr, using photon energy of 21.2 eV, linear horizontal polarization, with energy and momentum resolution better than 20 meV and 0.01 $Å^{-1}$ respectively. The bulk sample was mounted and cleaved in-situ inside the ultra-high vacuum (UHV) main chamber. The incident X-ray radiation was focused down to spot size of 20 microns and the photoemitted electrons were detected using an R4000 analyser equipped with a deflector. The energy versus momentum slices of the device along the high symmetry direction were obtained by orienting the detector in a $\overline{\Gamma}-\overline{\mathrm{M}}$ direction. Qualitative analysis from peak fitting was performed by fitting Voigt function by convoluting the gaussian peak representing the experiment resolution and the Lorentzian peak containing the quantitative information of the quasiparticle.

## *Interfacial thermal conductance measurement by FDTR*

We employed a home-built FDTR system (**fig. S3**) consisting of a 150 mW, 488 nm pump laser and an 80 mW, 532 nm probe laser (OBIS, Coherent). The pump beam was power-modulated using the reference output of a lock-in amplifier (HF2LI, Zurich Instruments), while the probe beam monitored the resulting surface temperature oscillations through thermoreflectance. Both beams were spatially overlapped and focused onto the sample through a 20× microscope objective. Precise beam overlap and focusing were ensured by maximizing the image sharpness of the sample surface acquired with a CCD camera (CS165CU, Thorlabs). Once aligned, the measurement location was kept fixed throughout each temperature- and current-dependent measurement sequence to minimize artifacts associated with beam realignment and spatial inhomogeneity.

The reflected probe signal was detected using a balanced photodetector (PDB425A, Thorlabs), and the differential signal was fed into the lock-in amplifier, which measured the in-phase ($V_{in}$) and out-of-phase ($V_{out}$) components at each modulation frequency. The thermoreflectance response ($\varphi_{probe}$) was defined as $\varphi_{probe} = \tan^{-1}\left(V_{out}/V_{in}\right)$. To account for parasitic phase delays from optical components, we performed a calibration by blocking the probe beam and measuring the corresponding background signal of the pump beam alone ($\varphi_{pump}$). The corrected thermal phase signal ($\varphi$) was then obtained as $\varphi = \varphi_{probe} - \varphi_{pump}$. **fig. S4** presents representative $\varphi$ data as a function of $f$ for the Au/$Bi_{89}Sb_{11}$ junction measured at 80 K. Eight consecutive frequency sweeps were performed under identical conditions. The relative standard deviation of the phase signal remained below 0.5% across the full frequency range, there by independent fitting of the eight phase spectra yielded a relative standard deviation of approximately 0.3% in the extracted $G_{[001]}$ values. These results confirm the excellent repeatability of the FDTR measurements.

To extract $G_{[001]}$ at the Au/$Bi_{1-x}Sb_x$ interface, the experimental thermal phase data were fitted using a multilayer thermal diffusion model (See Refs. (*10-13*) for details). Parameters independently measured or obtained from the literature were fixed during the fitting procedure, including the Au layer thickness (100 nm), the thermal conductivity and heat capacity of Au, and the heat capacity of $Bi_{1-x}Sb_x$ (*69*). The cross-plane thermal conductivity of $Bi_{1-x}Sb_x$ ($\kappa_{[001]}$) was determined independently by steady-state measurements (main text, **Fig. 2D**), while the in-plane values were estimated using the reported anisotropy ratio (*70*). The resulting $G_{[001]}$ values were subsequently analyzed as functions of $T$ and $j$ (main text, **Fig. 4** and **Fig. 6**).

A sensitivity analysis of the phase response to $G_{[001]}$ was performed for the Au/$Bi_{87}Sb_{13}$ sample at 91 K (**fig. S5**). The results confirmed that the phase retained appreciable sensitivity to $G_{[001]}$ across the full modulation frequency range, with particularly strong sensitivity at higher frequencies. The well-structured spectral response, combined with tightly constrained parameters, enabled reliable and accurate extraction of $G_{[001]}$.

For current injection experiments (**fig. S6**), electrical contacts were established using copper wires and silver epoxy, connecting the sample surface and the copper heat sink to a current source (Model 6221, Keithley). The voltage drop across the sample was measured using a nanovoltmeter (Model 2182A, Keithley) connected to contacts on the sample surface and side. The resulting current-voltage characteristics were linear over the investigated current range, confirming Ohmic contact. Possible artifacts associated with Joule heating were evaluated in **Note S1**.

## *FDTR Derivation of electron-electron thermal conductance within the eDMM*

We employed For electron-mediated interfacial thermal transport, the total thermal resistance can be represented as a series combination of interfacial electron-electron coupling ($G_{ee}$) and subsequent energy relaxation into the $Bi_{1-x}Sb_x$ bulk ($G_{bulk}$):

$$\frac{1}{G_{tot}} = \frac{1}{G_{ee}} + \frac{1}{G_{bulk}} \quad (5)$$

Because the bulk constitutes a large electronic and phononic reservoir, energy relaxation within the bulk is expected to be substantially faster than interfacial electron transmission, i.e., $G_{\mathrm{bulk}} \gg G_{\mathrm{ee}}$. Under this condition, $G_{\mathrm{tot}} \approx G_{\mathrm{ee}}$. Therefore, the measured $G_{[001]}$ is primarily determined by the electron-electron transmission channel.

Having established that $G_{\mathrm{tot}}$ is governed primarily by interfacial electron-electron transmission, we next evaluate $G_{\mathrm{ee}}$ within the framework of the eDMM.

Within the eDMM, $G_{\mathrm{ee}}$ between Au and electronic channel $i$ is given by

$$G_{\mathrm{ee}} = \frac{Z_{\mathrm{Au}} Z_i}{4\left(Z_{\mathrm{Au}} + Z_i\right)} \quad (6)$$

where $Z$ is the electronic thermal impedance,

$$Z = \frac{\pi^2 k_{\mathrm{B}}{}^2 T}{3} v_{\mathrm{F}} g\left(E_{\mathrm{F}}\right) \quad (7)$$

Here, $v_{\mathrm{F}}$ denotes the Fermi velocity and $g(E_{\mathrm{F}})$ is the electronic DOS at the Fermi level.

Since the electronic DOS of Au substantially exceeds that of all $Bi_{1-x}Sb_x$ electronic channels considered here, Eq. (6) reduces to

$$G_{\mathrm{ee}} \approx \frac{\pi^2 k_{\mathrm{B}}{}^2 T}{12} v_{\mathrm{F}} g\left(E_{\mathrm{F}}\right) \quad (8)$$

In the absence of TIS, interfacial thermal transport is mediated by bulk L-band electrons. For an anisotropic parabolic band, the dispersion can be expressed as

$$E(k) = E_{\mathrm{C(V)}} \pm \left( \frac{\hbar^2 k_x{}^2}{2m_x} + \frac{\hbar^2 k_y{}^2}{2m_y} + \frac{\hbar^2 k_z{}^2}{2m_z} \right) = E_{\mathrm{C(V)}} \pm \frac{\hbar^2 \tilde{k}^2}{2m^*_{\mathrm{DOS,\,3D}}} \quad (9)$$

where $m^*_{\mathrm{DOS,3D}} = \left(m_x m_y m_z\right)^{1/3}$ is the 3D DOS effective mass.

Accounting for the sixfold valley degeneracy, the DOS at the Fermi level is given by

$$g_{\text{bulk}}\left(E_{\text{F}}\right)=\frac{6m_{\text{DOS,3D}}^{*}{}^{3/2}}{\pi^{2}\hbar^{3}}\sqrt{2\left|E_{\text{C(V)}}-E_{\text{F}}\right|} \quad (10)$$

and the Fermi velocity is

$$v_{\text{F,bulk}}=\sqrt{\frac{2\left|E_{\text{C(V)}}-E_{\text{F}}\right|}{m_{\text{DOS,3D}}^{*}}} \quad (11)$$

Substituting Eqs. (10) and (11) into Eq. (8) yields

$$G_{\text{ee1}}\approx\frac{k_{\text{B}}{}^{2}T}{\hbar^{3}}m_{\text{DOS,3D}}^{*}\left|E_{\text{C(V)}}-E_{\text{F}}\right| \quad (12)$$

For the TIS channels, separate expressions are derived for the linear S1 branch and the parabolic S1’ and S2 pockets.

For the linear S1 branch, the dispersion is given by

$$E(k)=E_{0}\pm\hbar v_{\text{F}}k \quad (13)$$

The corresponding two-dimensional DOS at the Fermi level is

$$g_{\text{2D}}=\frac{\left|E_{0}-E_{\text{F}}\right|}{2\pi\hbar^{2}v_{\text{F}}{}^{2}} \quad (14)$$

Assuming that the spatial extent of the TIS wavefunctions is given by its de Broglie wavelength, $\lambda=hv_{\text{F}}/\left|E_{0}-E_{\text{F}}\right|$, the effective three-dimensional DOS becomes

$$g_{\text{S1}}=\frac{\left(E_{0}-E_{\text{F}}\right)^{2}}{4\pi^{2}\hbar^{3}v_{\text{F}}{}^{3}} \quad (15)$$

Using Eq. (15) in Eq. (8) gives

$$G_{\text{ee2,S1}}\approx\frac{k_{\text{B}}{}^{2}T}{48\hbar^{3}v_{\text{F}}{}^{2}}\left(E_{0}-E_{\text{F}}\right)^{2} \quad (16)$$

For the parabolic S1’ and S2 pockets, the dispersion can similarly be written as

$$E(k)=E_{\mathrm{C(V)}}\pm\left(\frac{\hbar^2 k_x{}^2}{2m_x}+\frac{\hbar^2 k_y{}^2}{2m_y}\right)=E_{\mathrm{C(V)}}\pm\frac{\hbar^2\tilde{k}^2}{2m^*_{\mathrm{DOS,2D}}} \quad (17)$$

where $m^*_{\mathrm{DOS,2D}}=\left(m_x m_y\right)^{1/2}$ is the 2D DOS effective mass.

The resulting DOS are given by

$$g_{\mathrm{S1'}}\left(E_{\mathrm{F}}\right)=\frac{3m^*_{\mathrm{DOS,3D}}{}^{3/2}}{\pi^2\hbar^3}\sqrt{2\left|E_{\mathrm{C}}-E_{\mathrm{F}}\right|} \quad (18) \text{ and } g_{\mathrm{S2}}\left(E_{\mathrm{F}}\right)=\frac{3m^*_{\mathrm{DOS,3D}}{}^{3/2}}{\pi^2\hbar^3}\sqrt{2\left|E_{\mathrm{V}}-E_{\mathrm{F}}\right|} \quad (19)$$

The Fermi velocity takes the form

$$v_{\mathrm{F,S1'}}=\sqrt{\frac{2\left|E_{\mathrm{C}}-E_{\mathrm{F}}\right|}{m^*_{\mathrm{DOS,2D}}}} \quad (20) \text{ and } v_{\mathrm{F,S2}}=\sqrt{\frac{2\left|E_{\mathrm{V}}-E_{\mathrm{F}}\right|}{m^*_{\mathrm{DOS,2D}}}} \quad (21)$$

Applying these expressions to Eq. (8) leads to

$$G_{\mathrm{ee2,S1'}}\approx\frac{k_{\mathrm{B}}{}^2 T}{2\hbar^3}m^*_{\mathrm{DOS,2D}}\left|E_{\mathrm{C}}-E_{\mathrm{F}}\right| \quad (22) \text{ and } G_{\mathrm{ee2,S2}}\approx\frac{k_{\mathrm{B}}{}^2 T}{2\hbar^3}m^*_{\mathrm{DOS,2D}}\left|E_{\mathrm{V}}-E_{\mathrm{F}}\right| \quad (23)$$

## Acknowledgements

The authors dedicate this work to the memory of Joseph P. Heremans. His scientific leadership, encouragement, and unwavering support were instrumental to this research and continue to inspire us.
M.Y.K., J.S.K. and J.P.H. acknowledge support from the U.S. Office of Naval Research MURI program, "Extraordinary electronic switching of thermal transport" (N00014-21-1-2377). M.Y.K. acknowledges support from the Korea Institute of Science and Technology (KIST) (Project No. 26Z9052). J.H. and H.J. acknowledge support from the National Research Foundation of Korea (NRF), funded by the Ministry of Science and ICT (MSIT) (RS-2024-00345022 and NRF-2022M3C1A3091988). J.K. acknowledges financial support from the U.S. Department Office of Energy, Office of Science, Office of Basic Energy Sciences, of the U.S. Department of Energy through award No. DE-SC002549 for S.A.E. support. ARPES data were obtained from MAESTRO beamline in the Advanced Light Source, which is a DOE Office of Science User Facility under contract No. DE-AC02-05CH11231.

## Conflicts of interest

There are no conflicts to declare.

## Data Availability Statement

All processed data are present in the main text and/or the Supplementary Materials. Raw data are available from the corresponding author upon reasonable request.

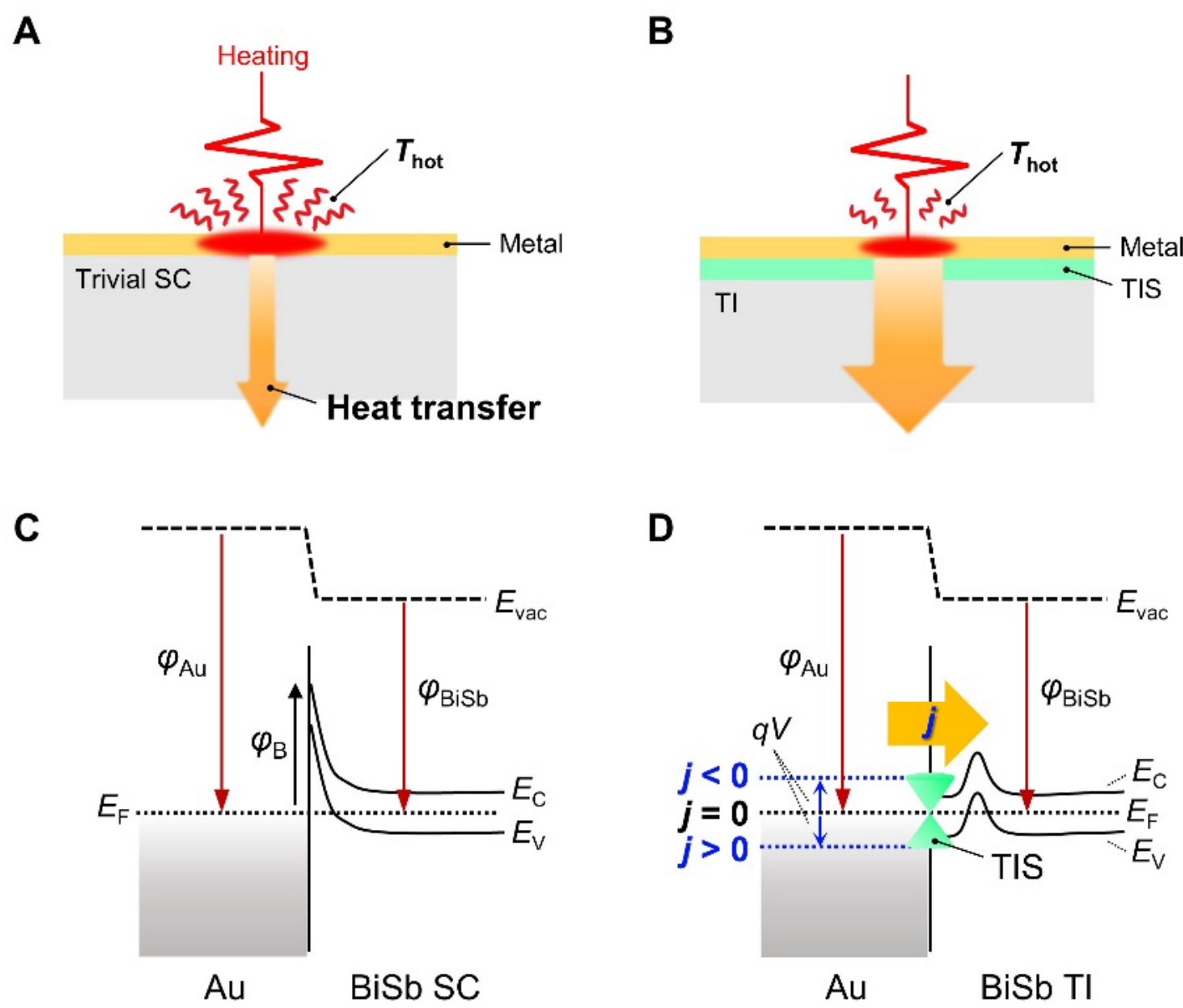

**Fig. 1. Concept of TIS-mediated interfacial thermal conduction and electrical modulation of interfacial thermal conductance.** (**A**, **B**) Schematic illustrations of heat transfer through (**A**) a metal-trivial semiconductor (SC) junction and (**B**) a metal-TI junction (*G*: Interfacial thermal conductance). In the metal-TI junction, TIS emerge at the interface and provide an additional electronic pathway that contributes to interfacial heat transfer. (**C**, **D**) Schematic band diagrams of Au/$Bi_{1-x}Sb_x$ junctions in the trivial SC (**C**) and TI (**D**) phases. Key features include the Schottky barrier height ($\varphi_B$), the work function of Au ($\varphi_{Au} \approx 5.1$ eV), and that of $Bi_{1-x}Sb_x$ ($\varphi_{BiSb} \approx 4.5$ eV) (*71, 72*). Also shown are the vacuum level $E_{vac}$, conduction-band edge $E_C$, valence-band edge $E_V$, and Fermi level $E_F$. In the TI phase, the TIS emerge near the interface with a high density of states (DOS), resulting in a richer electronic spectrum in the vicinity of the interface. An applied current density ($j$) across the junction induces an electrochemical potential difference at the interface by an amount of $qV$ ($q$: elementary charge; $V$: junction voltage).

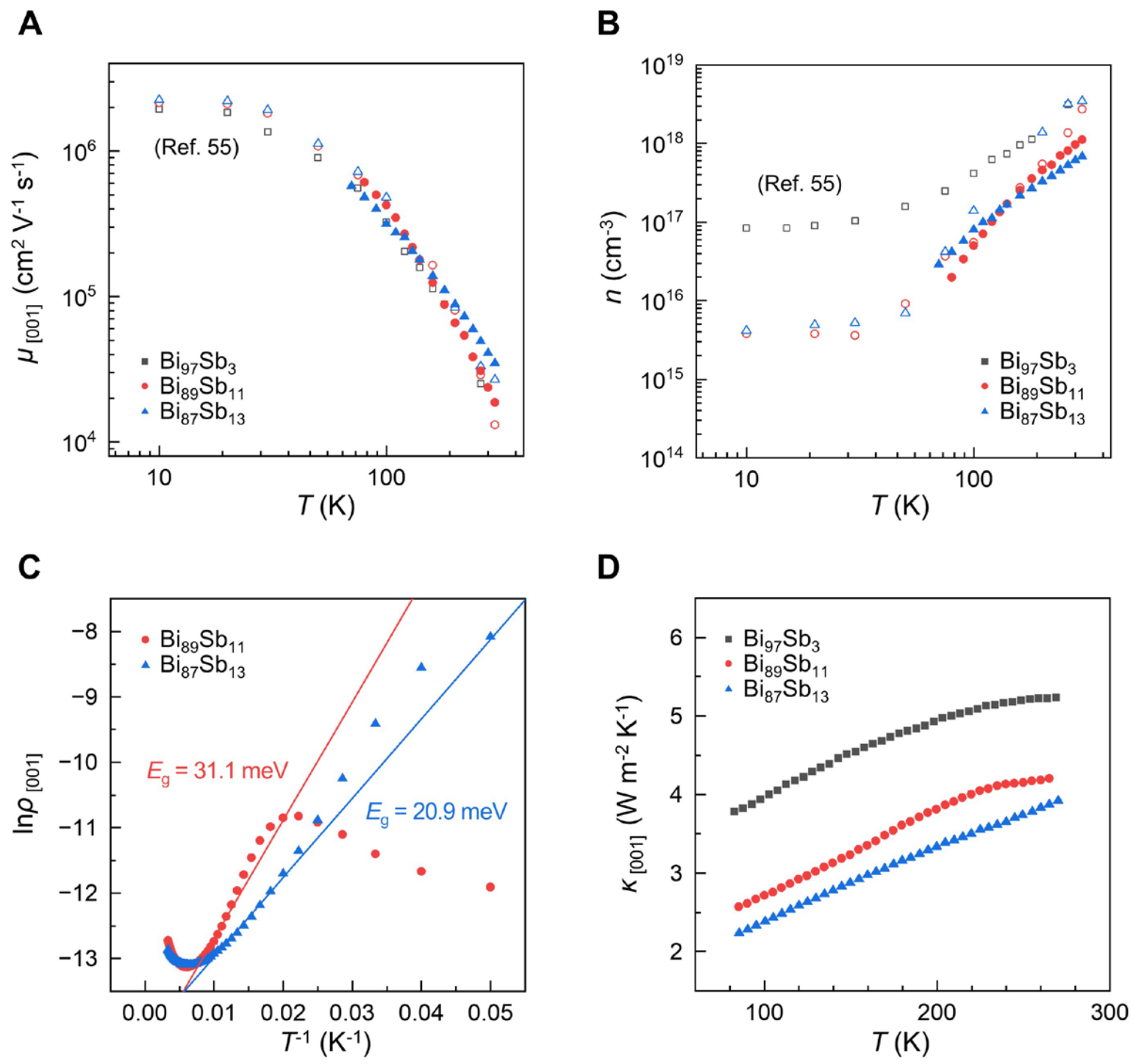

**Fig. 2. Electrical and thermal transport characterization of high-quality $Bi_{1-x}Sb_x$ single crystals**. (**A**) $T$-dependent carrier mobility $\mu_{[001]}$ and (**B**) carrier concentration $n$ of $Bi_{97}Sb_3$, $Bi_{89}Sb_{11}$, and $Bi_{87}Sb_{13}$. Filled symbols represent measurements performed in the present work, whereas open symbols denote data reproduced from Ref. (*55*). The high mobilities and distinct low-temperature carrier densities reflect the semimetallic character of $Bi_{97}Sb_3$ and the insulating bulk character of $Bi_{89}Sb_{11}$ and $Bi_{87}Sb_{13}$. (**C**) Arrhenius plots used to extract the bulk band gaps, yielding $E_g \approx 31.1$ meV for $Bi_{89}Sb_{11}$ and 20.9 meV for $Bi_{87}Sb_{13}$ (non-monotonic with $x$; see main text). (**D**) Thermal conductivity $\kappa_{[001]}$ as a function of $T$, decreasing in magnitude with increasing $x$, consistent with enhanced phonon scattering from Sb incorporation.

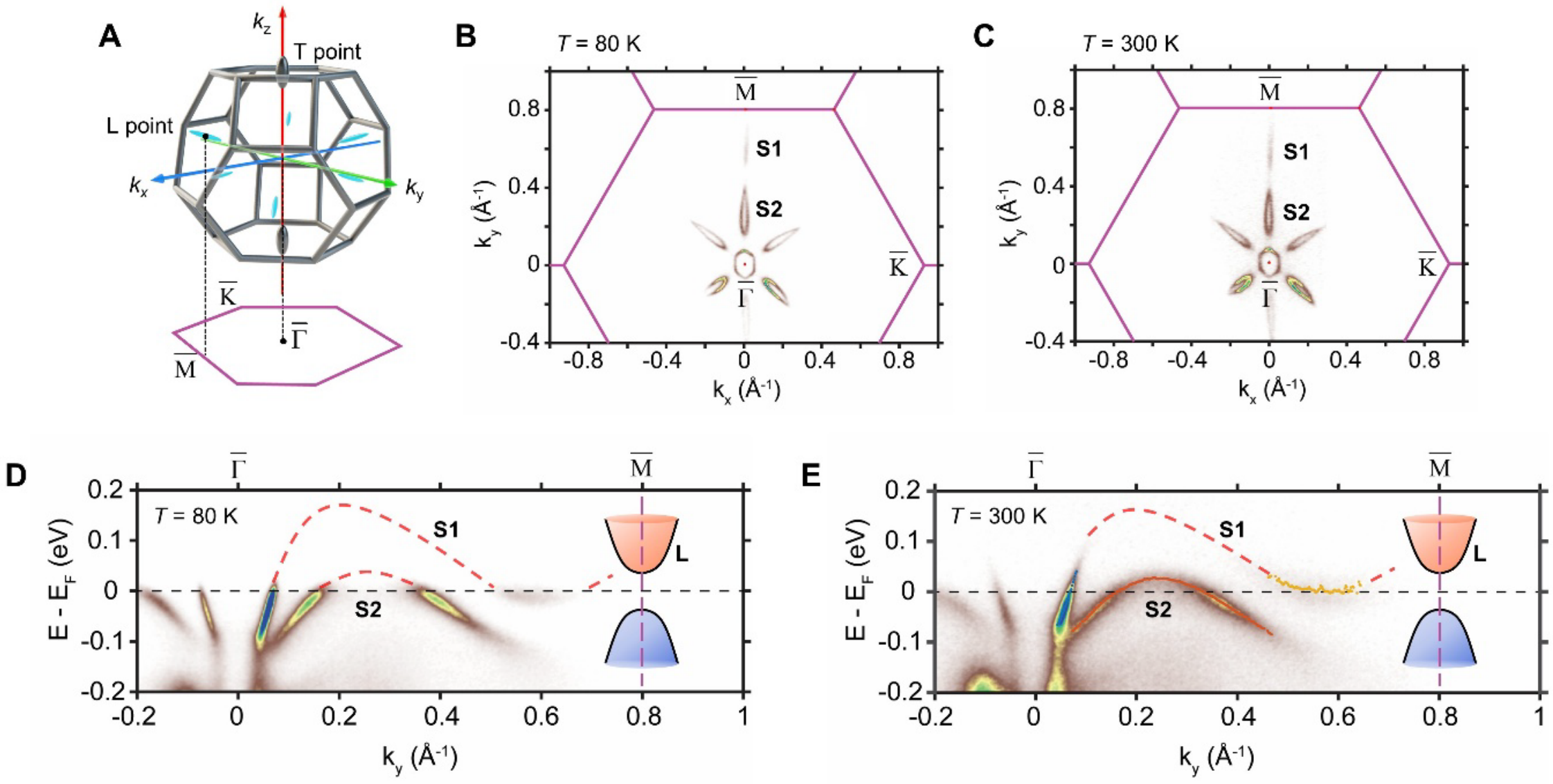

**Fig. 3. ARPES characterization of TSS in $Bi_{89}Sb_{11}$.** (**A**) Schematic of the bulk and projected (001) surface Brillouin zones of elemental Bi. The bulk Fermi surface contains electron and hole pockets near the L and T points, respectively. (**B**, **C**) Fermi-surface maps of $Bi_{89}Sb_{11}$ (001) at (**B**) 80 K and (**C**) 300 K, respectively. (**D**, **E**) Surface-state band dispersions of $Bi_{89}Sb_{11}$ (001) along the $\bar{\Gamma}-\bar{M}$ direction at (**D**) 80 K and (**E**) 300 K. Two surface bands, denoted S1 and S2, are observed. S1 evolves into an electron pocket near $\bar{M}$ (hereafter S1'), whereas S2 forms a hole pocket around $\bar{\Gamma}$. Together, the surface-state dispersions cross the Fermi level five times, confirming the topologically non-trivial character of the system. The bulk L-bands near $\bar{M}$ are indicated schematically, and the red dashed lines are guides to the eye for the S1 and S2 branches.

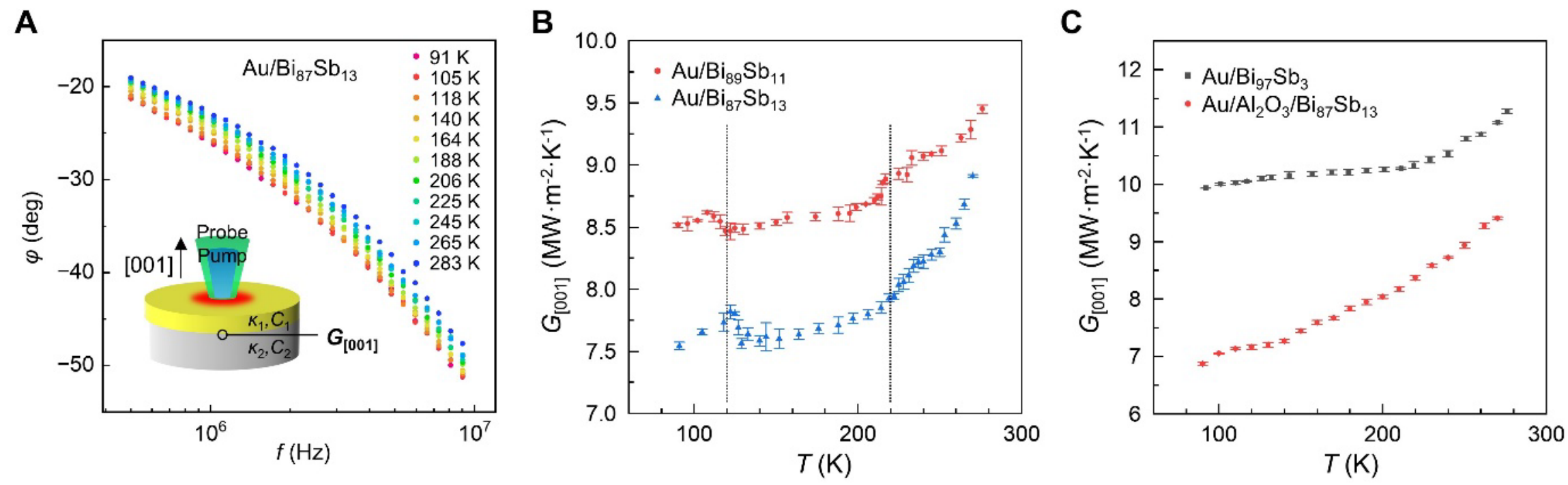

**Fig. 4. Temperature-dependent interfacial thermal conductance in Au/Bi$_{1\text{-}x}$Sb$_x$ junctions**. (**A**) Representative FDTR $\varphi$ traces of Au/$Bi_{87}Sb_{13}$ sample as a function of $f$ at various temperatures, from which $G_{[001]}$ is extracted by fitting to a thermal diffusion model that includes the thermal conductivity $\kappa_i$ and heat capacity $C_i$ of $i$-th layer. (**B**) $T$-dependent $G_{[001]}$ for two TI samples with $x$ of 11±1% and 13±1%. $G_{[001]}$ shows a non-monotonic $T$ dependence, being ~2% larger at $T < 120$ K than in the 120 K $< T <$ 220 K range, and rising again above 220 K. The relative standard deviation in $G_{[001]}$ is below 1%, providing sufficient sensitivity to resolve the observed changes in $G_{[001]}$. (**C**) $T$-dependence of $G_{[001]}$ in two control samples: Au/$Bi_{97}Sb_3$ (semimetal) and Au/$Al_2O_3$/$Bi_{87}Sb_{13}$ (electronically decoupled). Both show smooth, monotonic increases with $T$, in contrast to the non-monotonic trends observed in the Au/TI samples (**B**).

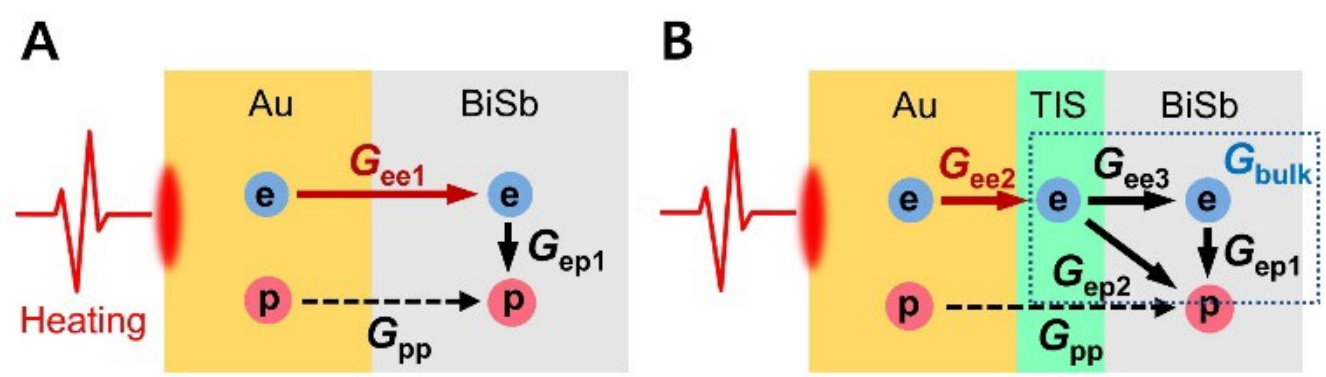

**Fig. 5. Schematic of the dominant thermal transport pathways across the Au-$Bi_{1-x}Sb_x$ interface within the framework of eDMM.** (**A**) In the absence of TIS, heat flows via electron-electron coupling ($G_{ee1}$) between Au and $Bi_{1-x}Sb_x$ electrons, followed by energy transfer to $Bi_{1-x}Sb_x$ phonon reservoir via bulk electron-phonon coupling ($G_{ep1}$). (**B**) In the presence of TIS, heat conduction occurs through electron-electron coupling ($G_{ee2}$) between Au electrons and TIS, followed by energy relaxation via either an indirect pathway ($G_{ee3} \rightarrow G_{ep1}$) or a direct electron-phonon coupling associated with TIS ($G_{ep2}$), collectively represented by $G_{bulk}$. Phonon-phonon coupling ($G_{pp}$, dashed) may also contribute to interfacial thermal transport in both cases.

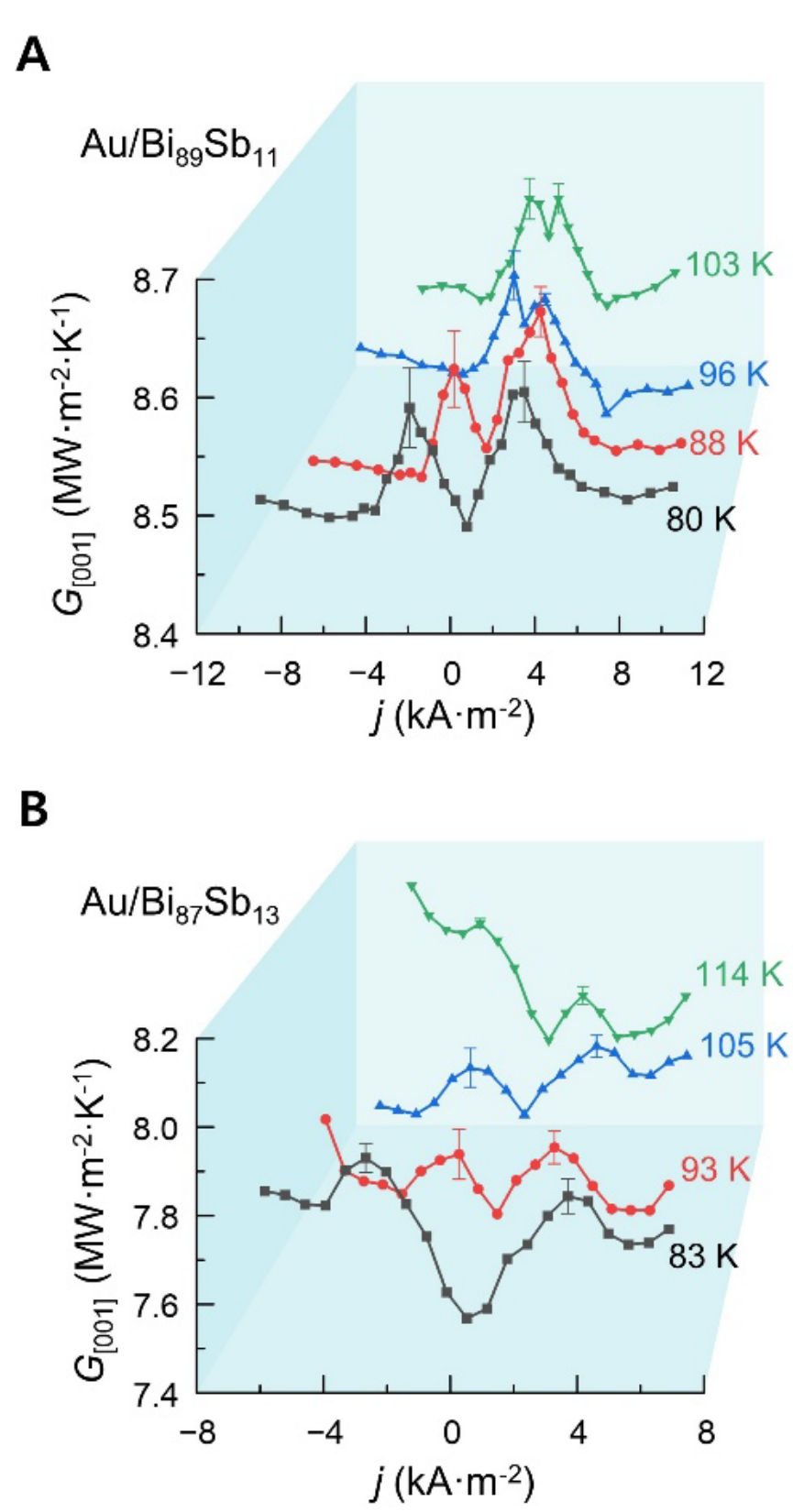

**Fig. 6. Electrically tunable interfacial thermal conductance in Au/$Bi_{89}Sb_{11}$ and Au/$Bi_{87}Sb_{13}$ junctions**. (**A**, **B**) $G_{[001]}$ as a function of applied bias current density $j$ for (**A**) Au/$Bi_{89}Sb_{11}$ and (**B**) Au/$Bi_{87}Sb_{13}$ samples below 120 K. A clear peak is observed in each case, followed by a sharp decline at higher bias. The low-bias regime is dominated by TIS-mediated transport, whereas the conductance suppression at higher $|j|$ reflects the onset of L-band activation via quasi-Fermi-level-induced tunneling.

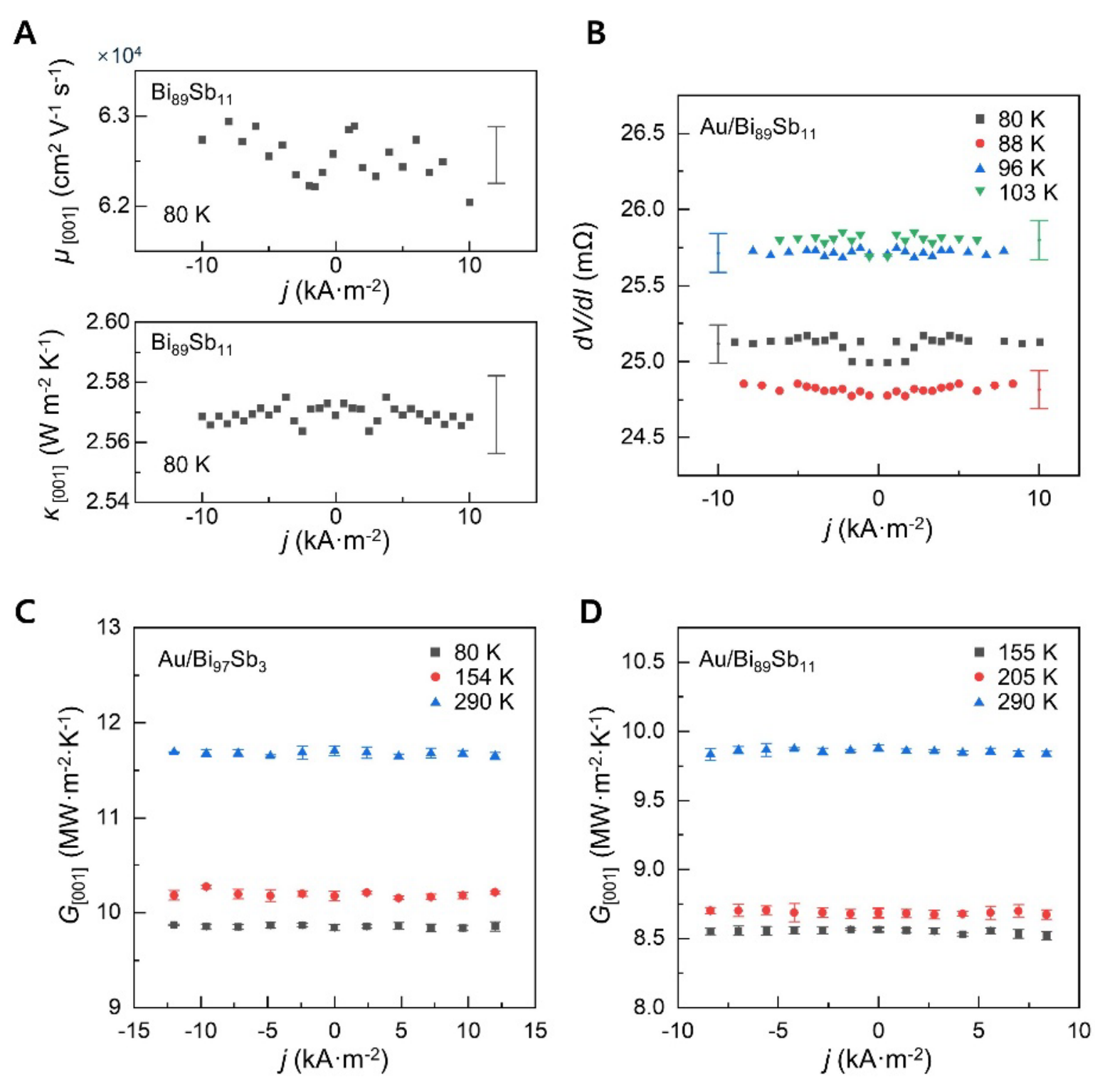

**Fig. 7. Absence of current-induced $G_{[001]}$ modulation outside the TIS-dominated transport regime.** (**A**) $j$ dependence of $\mu_{[001]}$ (top) and $\kappa_{[001]}$ (bottom) in $Bi_{89}Sb_{11}$ at 80 K. Both remain nearly constant over the entire current range, indicating that the bulk carrier and heat transport are unaffected by current injection. (**B**) Differential resistance $dV/dI$ of the Au/$Bi_{89}Sb_{11}$ junction as a function of $j$ between 80 and 103 K, showing no appreciable variation across the investigated current range. (**C**) $j$ dependence of $G_{[001]}$ in Au/$Bi_{89}Sb_{11}$ at elevated temperatures (155, 205, and 290 K). $G_{[001]}$ remains nearly constant across the full current range, consistent with the diminished role of TIS occupation at elevated temperatures. (**D**) $j$ dependence of $G_{[001]}$ in the control Au/$Bi_{97}Sb_3$ sample at 80, 154, and 290 K. The absence of current sensitivity in this TIS-free composition confirms that the current-induced modulation of $G_{[001]}$ requires the presence of TIS.

# Supplementary Materials

## *Supplementary Note*

### Note S1. Joule heating

Concerning the possibility that Joule heating of the sample at high current levels could potentially raise the sample temperature, there are 3 points to be made. First, the Joule heat to be dissipated is quite modest, < 0.12 mW, for the large size of the sample. Second, the strong thermal coupling to the copper sink over a surface of tens of $mm^2$ provides efficient heat dissipation. Third, any Joule heating effect would result in a monotonic quadratic shift of the curves in **Fig. 6** (**A** and **B**) and cannot give rise to a maximum, as observed. Therefore, significant thermal drift during FDTR measurements is unlikely under the experimental conditions employed in this study.

*Supplementary Figures*

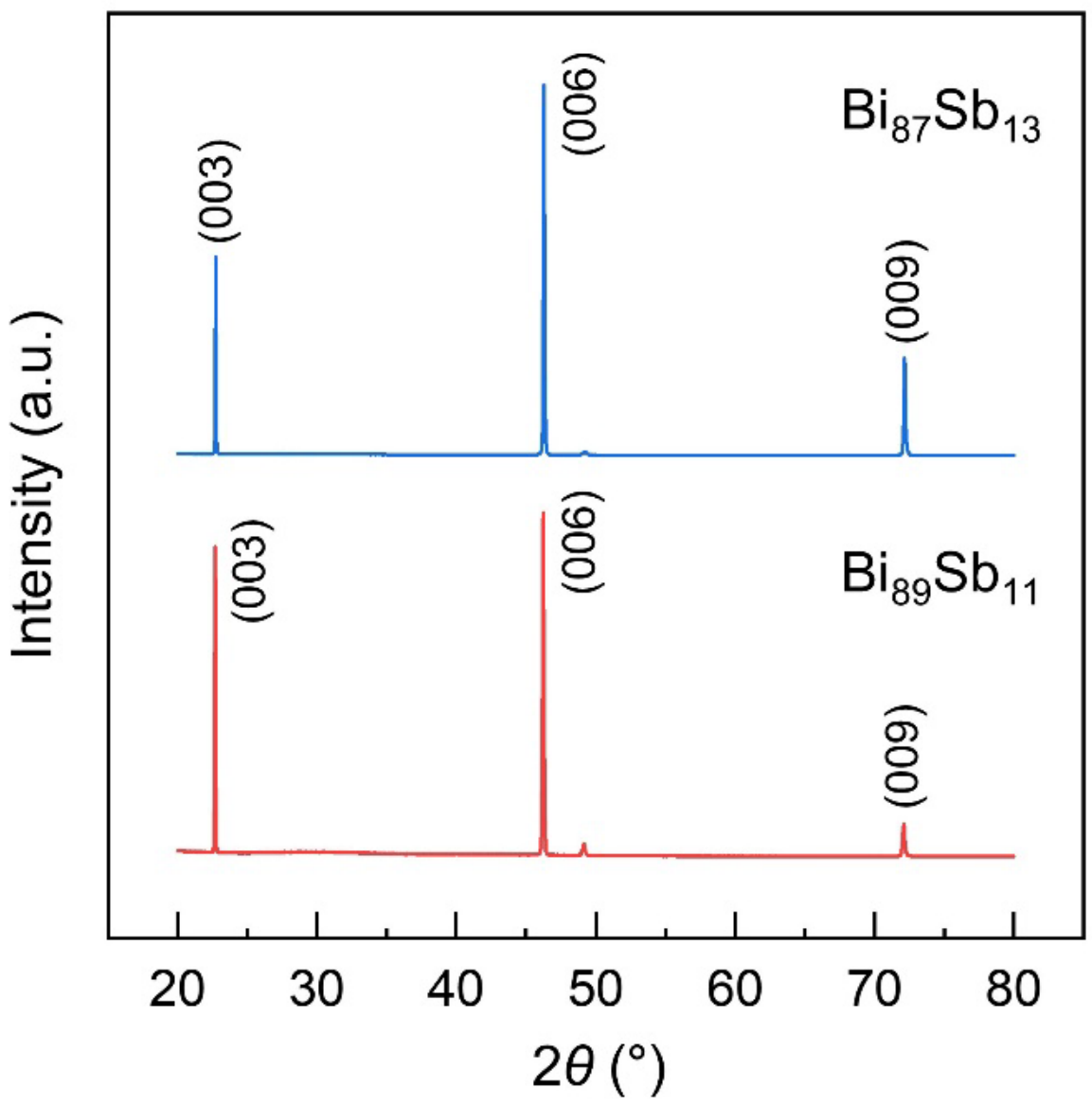

**Fig. S1.**

X-ray diffraction (XRD) patterns of $Bi_{89}Sb_{11}$ and $Bi_{87}Sb_{13}$ measured on the cleaved surface. Only (00*l*) reflections are observed, confirming that the exposed surface is oriented along the [001] crystallographic plane.

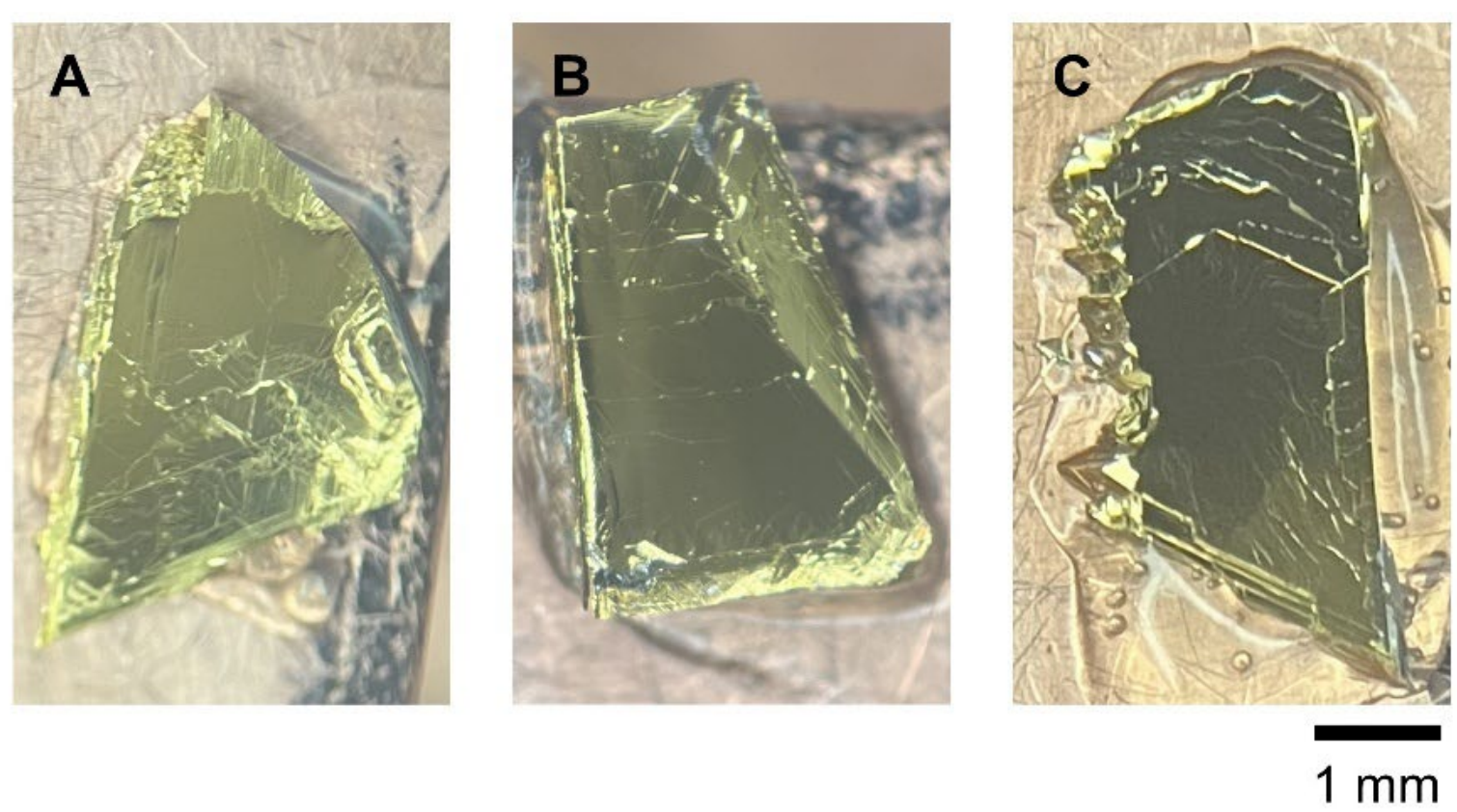

**Fig. S2.**

Optical images of $Bi_{1-x}Sb_x$ single crystals after deposition of a 100 nm-thick Au layer on the cleaved surfaces: (**A**) $Bi_{97}Sb_3$, (**B**) $Bi_{89}Sb_{11}$, and (**C**) $Bi_{87}Sb_{13}$.

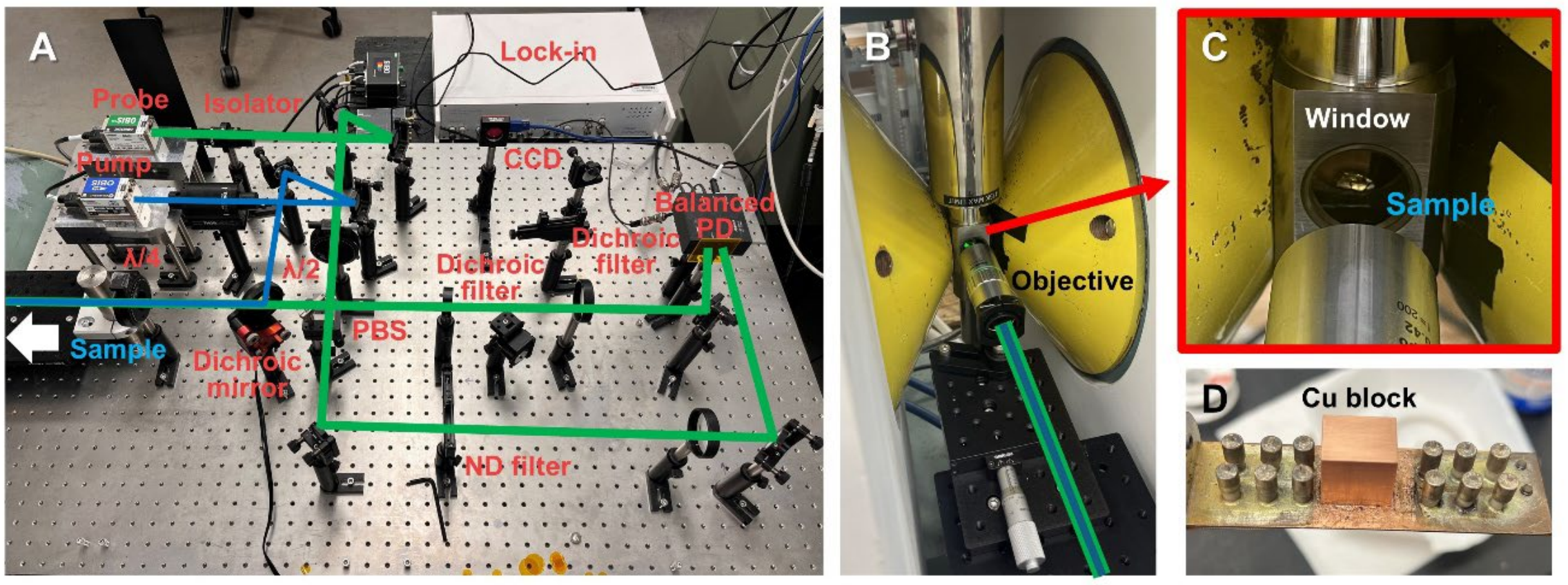

**Fig. S3.**

Photographs of the FDTR setup. (**A**) Top view of the full optical layout, including optical isolators, dichroic filters, PBS, λ/2 and λ/4 waveplates, ND filter, and a balanced photodetector (PD). (**B**) Side view of the cryogenic vacuum chamber, showing the 20× objective lens aligned to the sample inside. The pump and probe beams are spatially overlapped and focused onto the sample through the objective. (**C**) Zoomed-in view through the optical window, showing the Au-coated $Bi_{1-x}Sb_x$ sample mounted inside the chamber. (**D**) The Cu heat sink (Cu block) used as a thermal anchor for the sample.

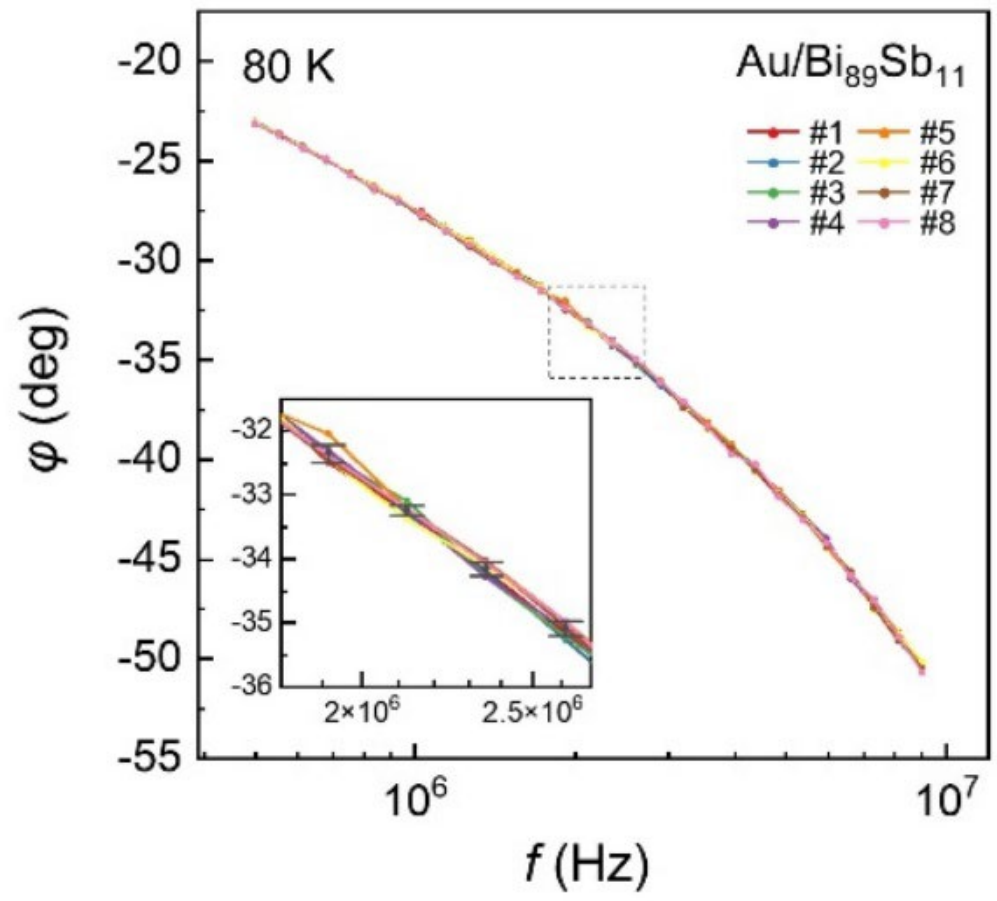

**Fig. S4.**

Thermal phase response $\varphi$ as a function of modulation frequency $f$ for the $Au/Bi_{89}Sb_{11}$ sample at 80 K. Eight independent frequency sweeps (#1-#8) show high reproducibility. The inset provides an expanded view of the dashed region, showing phase variations below 0.5%

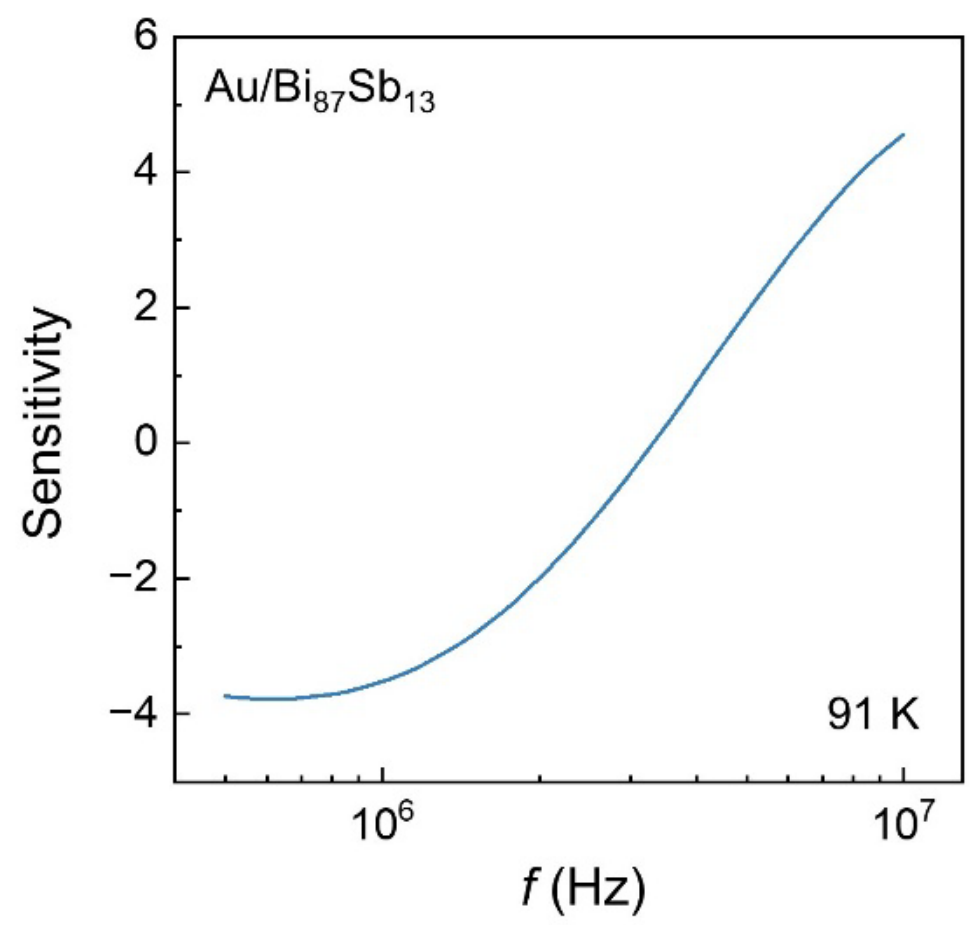

**Fig. S5.**

Sensitivity coefficient of the thermal phase response $\varphi$ with respect to $G_{[001]}$ for the Au/$Bi_{87}Sb_{13}$ sample at 91 K as a function of modulation frequency *f*. The phase response shows appreciable sensitivity to $G_{[001]}$ across the full frequency range, with particularly storng sensitivity at higher modulation frequencies.

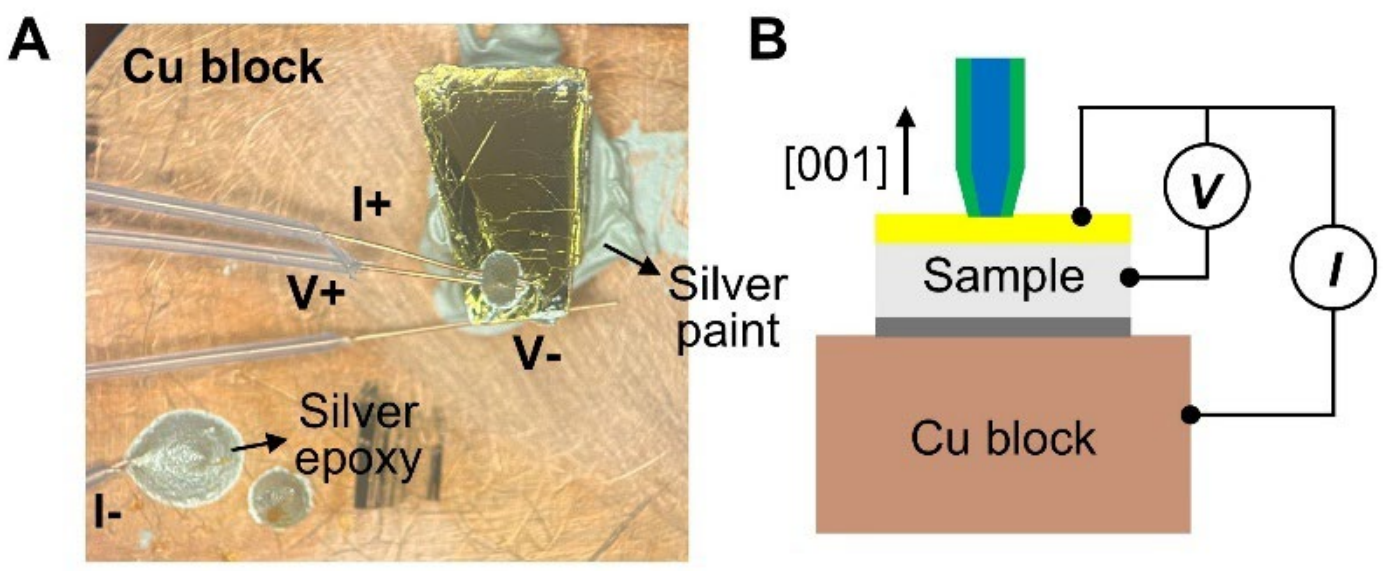

**Fig. S6.**

Electrical contact configuration for current injection experiments. (**A**) Optical image of an Au/$Bi_{1-x}Sb_x$ sample mounted on a Cu block, showing current (*I*+/*I*-) and voltage (*V*+/*V*-) leads connected to the sample and Cu block using silver epoxy and silver paint. (**B**) Schematic diagram of the measurement setup, where current is injected through the *I*+ and *I*- contacts, and voltage is measured between the *V*+ contact on the sample surface and the *V*- contact on the side sample.

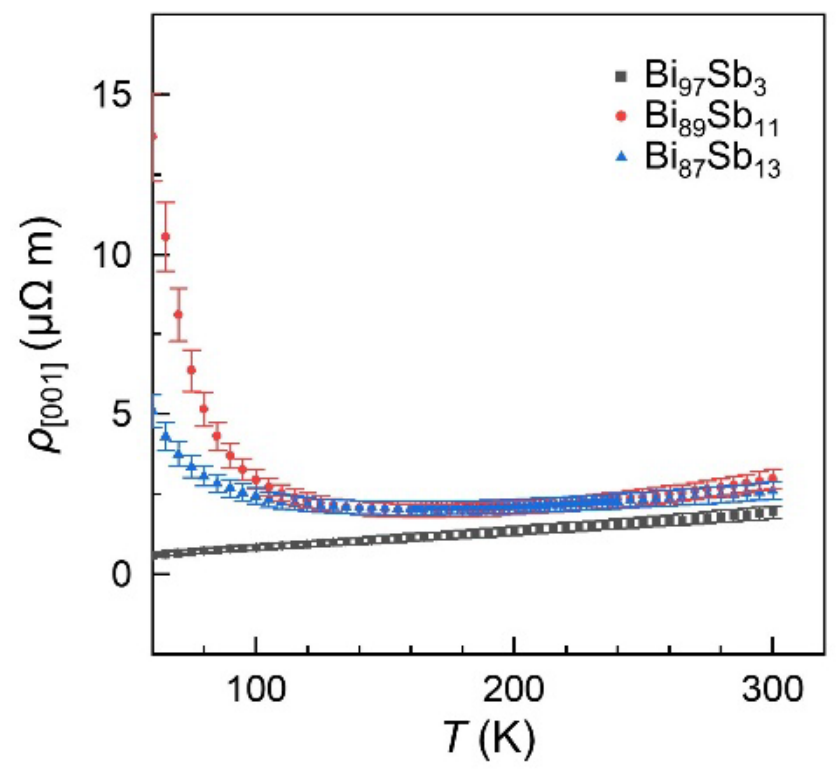

**Fig. S7.**

$T$-dependent resistivity $\rho_{[001]}$ of $Bi_{97}Sb_3$, $Bi_{89}Sb_{11}$, and $Bi_{87}Sb_{13}$. $Bi_{97}Sb_3$ behaves metallically, while $Bi_{89}Sb_{11}$ and $Bi_{87}Sb_{13}$ show toward more metallic transport behavior with increasing $T$.

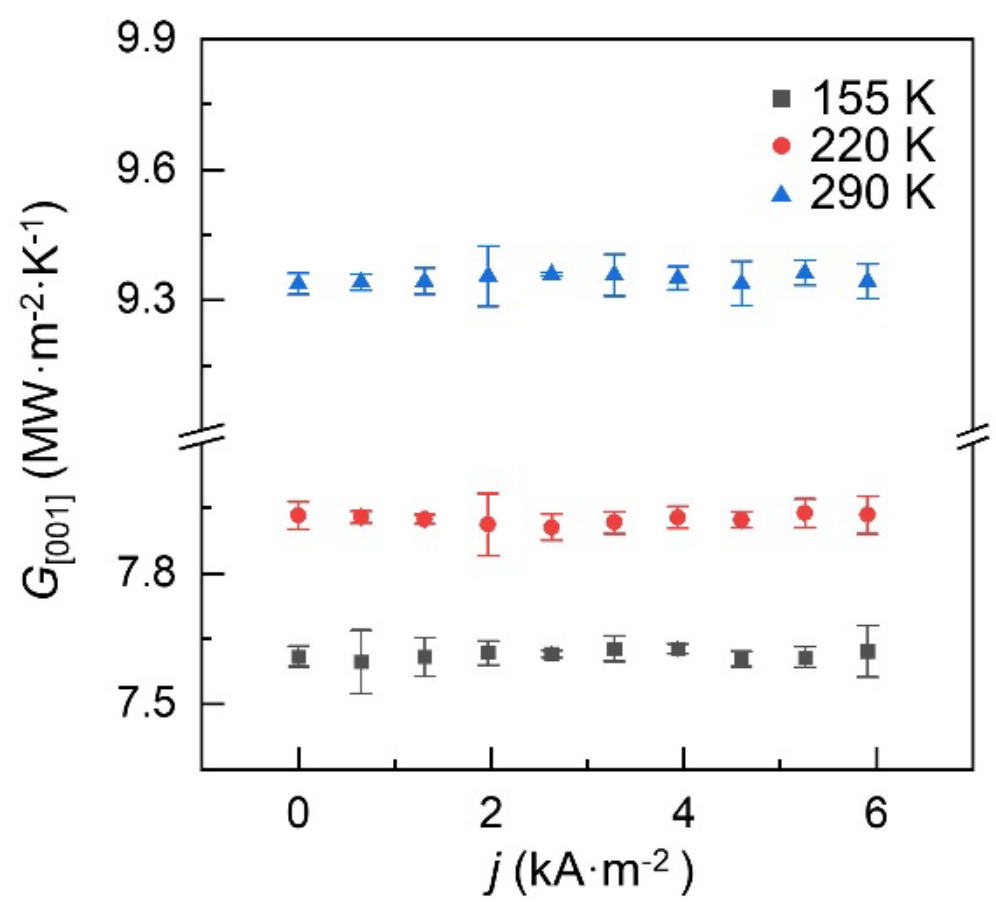

**Fig. S8.**

$j$-dependence of $G_{[001]}$ in Au/$Bi_{87}Sb_{13}$ at high temperatures (155, 220, and 290 K). $G_{[001]}$ remains nearly constant across the full current range, consistent with the reduced influence of TIS-mediated transport at these temperatures.

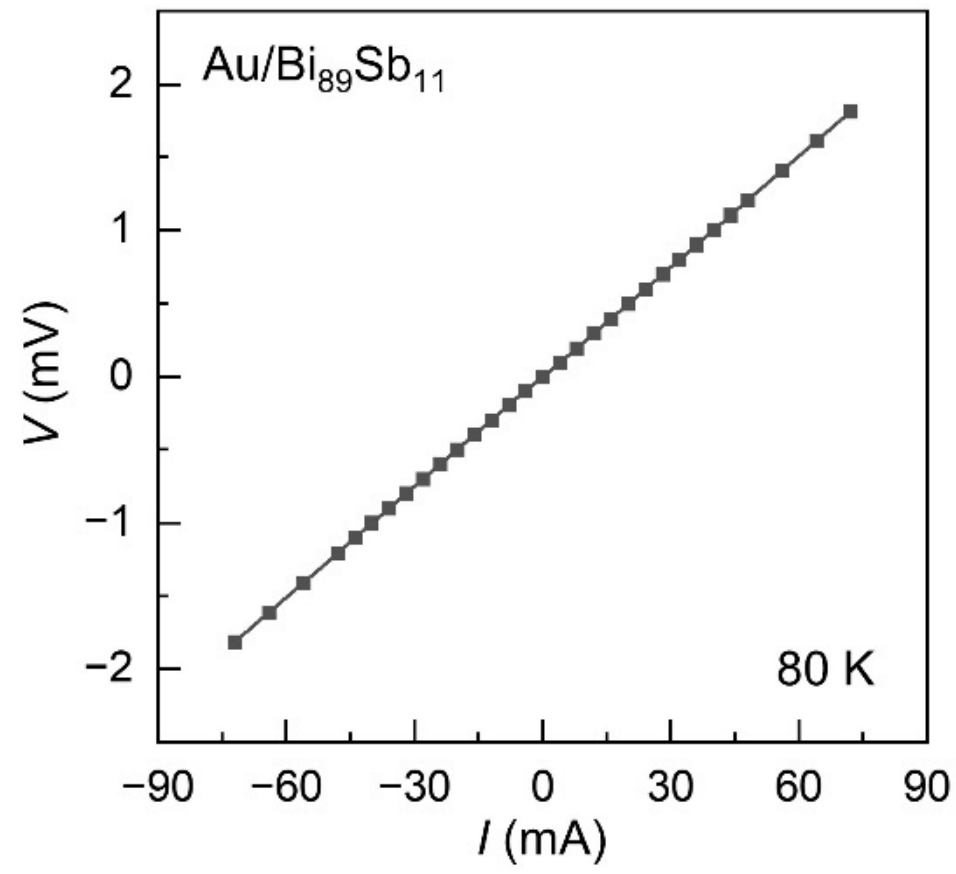

**Fig. S9.**

Current-voltage (*I*-*V*) characteristics of the Au/$Bi_{89}Sb_{11}$ sample at 80 K. The linear *I*-*V* relationship confirms Ohmic contact over the investigated current range.

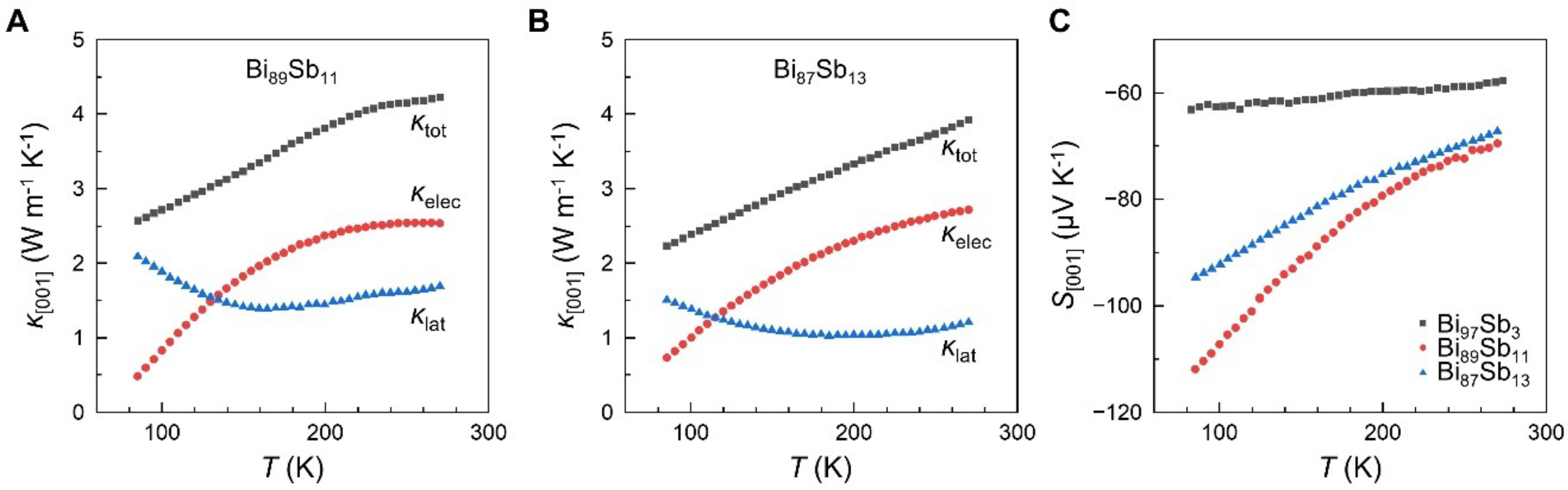

**Fig. S10.**

Cross-plane bulk thermal transport properties of $Bi_{1-x}Sb_x$ single crystals. (**A**, **B**) $T$-dependent thermal conductivity $\kappa_{[001]}$ along the [001] axis, decomposed into lattice ($\kappa_{lat}$) and electronic ($\kappa_{elec}$) contributions for (**A**) $Bi_{89}Sb_{11}$ and (**B**) $Bi_{87}Sb_{13}$. The electronic contribution $\kappa_{elec}$ was estimated from the electrical resistivity using the Wiedemann-Franz law, and $\kappa_{lat}$ was obtained by subtraction. (**C**) $T$-dependent Seebeck coefficient $S_{[001]}$ measured along the [001] axis.